\documentclass[letterpaper]{article}

\pdfoutput=1
\usepackage{jheppub}
\usepackage{graphicx, caption, subcaption}  
\usepackage{dcolumn}   
\usepackage{amsmath, amssymb}        
\usepackage{color,latexsym, array,multirow,  verbatim, enumerate,cancel}

\def\issue(#1,#2,#3){{\bf #1}, #2 (#3)}

\catcode`\@=11
\def\lsim{\mathrel{\mathpalette\@versim<}}
\def\gsim{\mathrel{\mathpalette\@versim>}}
\def\@versim#1#2{\vcenter{\offinterlineskip
\ialign{$\m@th#1\hfil##\hfil$\crcr#2\crcr\sim\crcr } }}
\catcode`\@=12

\catcode`@=12

\newcommand{\met}{$\cancel E_T$}

\newcommand{\newc}{\newcommand}
\newc{\wt}{\widetilde}
\newc{\ra}{\rightarrow}
\usepackage{multicol}
\def\beq {\begin{equation}}
\def\eeq {\end{equation}}
\def\bi {\begin{itemize}}
\def\ei {\end{itemize}}
\def\bea {\begin{eqnarray}}
\def\eea {\end{eqnarray}}
\def \PMET{\rm p{\!\!\!/}_T}

\def \met{\rm E{\!\!\!/}_T}

\newcommand{\br}{\begin{eqnarray}}
\newcommand{\er}{\end{eqnarray}}
\newcommand{\be}{\begin{equation}}
\newcommand{\ee}{\end{equation}}

\newcommand{\ch}{\widetilde \chi^\pm}

\def \chonep {{\wt\chi_1^+}}

\def \ch2p {{\wt\chi_2^+}}

\def \ch2m {{\wt\chi_2^-}}

\def \chonepm{{\wt\chi_1}^{\pm}}

\newc{\dmchi}{\Delta m_{\wt\chi}}
\def \chtwop {{\wt\chi_2^+}}

\def \chtwopm{{\wt\chi_2}^{\pm}}

\def \lspone{\wt\chi_1^0}

\def \lsptwo{\wt\chi_2^0}

\def \lspthree{\wt\chi_3^0}

\def \lspfour{\wt\chi_4^0}

\def\issue(#1,#2,#3){{\bf #1}, #2 (#3)}

\hypersetup{
   colorlinks=true,       
   linkcolor=blue,        
   citecolor=red,         
   filecolor=magenta,      
   }

\title{The invisible decay of Higgs boson in the context of a thermal and non-thermal relic in MSSM}

\author[a]{Rahool Kumar Barman,}
\author[b]{Genevieve B\'elanger,}
\author[a]{Biplob Bhattacherjee,}
\author[a]{Rohini Godbole,}
\author[a,e]{Gaurav Mendiratta,}
\author[c,d]{Dipan Sengupta}

\affiliation[a]{Center for High Energy Physics, 
Indian Institute of Science, Bangalore, India
}
\affiliation[b]{LAPTh, Universit\'e Savoie Mont Blanc, CNRS, B.P. 110,  
F-74941 Annecy Cedex, France}
\affiliation[c]{Department of Physics and Astronomy, Michigan State University, 
567 Wilson Road, \\ East Lansing, Michigan, USA}
\affiliation[d]{Laboratoire de Physique Subatomique et de Cosmologie, Universit\'e Grenoble-Alpes,\\ 
CNRS/IN2P3, 53 Avenue Des Martyrs, F-38026 Grenoble,France}
\affiliation[e]{Salk Institute for Biological Studies, 10010 N Torrey Pines Rd, \\ La Jolla, California, USA 92037}
\emailAdd{rahoolkbarman@chep.iisc.ernet.in}
\emailAdd{belanger@lapth.cnrs.fr}
\emailAdd{biplob@chep.iisc.ernet.in}
\emailAdd{rohini@cts.iisc.ernet.in}
\emailAdd{gauravm.137@gmail.com}
\emailAdd{dipan@pa.msu.edu}

\date{\today}

\abstract{We study the decay of 125 GeV Higgs boson to light LSP 
neutralino in the phenomenological minimal supersymmetric standard model 
in the context of collider searches and astrophysical experiments. We 
consider the parameter space for light neutralinos that can be probed 
via the invisible Higgs decays and higgsino searches at the ILC. We 
consider the cases where the light neutralino is compatible with the 
observed relic density or where the thermal relic is over-abundant, 
pointing to non-standard cosmology. In the former case, when the 
neutralino properties give rise to under-abundant relic density, the 
correct amount of relic abundance is assumed to be guaranteed by either 
additional DM particles or by non-thermal cosmology. We contrast these 
different cases. We assess what astrophysical measurements can be made, in 
addition to the measurements made at the ILC, which can provide a clue to 
the nature of the light neutralino. We find that a number of experiments, 
including Xenon-nT, PICO-250, LZ in conjunction with measurements made at 
the ILC on invisible Higgs width can pin down the nature of this 
neutralino, along with its cosmological implications. Additionally, we 
also point out potential LHC signatures that could  be complementary in 
this region of parameter space. 
}

\begin{document}
\maketitle

\section{Introduction}\label{sec:intro}
Particle physics today is at a juncture where all predicted particles 
within the Standard Model (SM) have been observed at particle colliders 
while no particles beyond the SM (BSM) have been detected. We have 
discovered a Higgs boson consistent with the properties of the SM 
\cite{Aad:2012tfa,Chatrchyan:2012xdj}. The current data, however still 
leaves enough space for the Higgs to have 
non-standard decays \cite{Khachatryan:2016vau}. One such possibility is 
that the Higgs acts as a portal to BSM physics at the electroweak scale. 
An exciting 
prospect in this regard is to consider the Higgs decaying to a pair of 
invisible particles in a BSM theory.  
Such invisible particles, 
if stable at the time scale of the universe, could also be the dark matter 
(DM) particle.  Prospects for the discovery of an invisible branching 
ratio of the Higgs at the LHC have been explored  in a number of studies 
e.g \cite{Gunion:1993jf,Choudhury:1993hv,Eboli:2000ze,Godbole:2003it, 
Davoudiasl:2004aj,Accomando:2006ga,Bai:2011wz,Djouadi:2012zc,Ghosh:2012ep,Bernaciak:2014pna}. 
In fact, both CMS and ATLAS have looked, in both run I and run II of LHC, 
for such invisibly decaying Higgs through its inclusive production in the 
gluon fusion, in the vector boson fusion mode, as well as in the 
associated production of a Higgs with a $Z$ boson. CMS has analysed the 
data corresponding to $5.1,~19.7\text{ and}~2.3~{\rm fb^{-1}}$ of 
integrated luminosity, collected at $\sqrt{s} = 7,~8~{\rm and}~13~{\rm 
TeV}$, respectively, and has obtained an upper limit on the invisible 
branching fraction $\sim 24\%$ at $95\%$ 
C.L.~\cite{Chatrchyan:2014tja,CMS-PAS-HIG-16-016,Khachatryan:2016whc}. 
ATLAS has also searched for the invisible decay of the Higgs boson, 
produced via the associated production of the Higgs with a $Z$ and via 
vector boson fusion\cite{Aad:2015pla,Aad:2015txa}. The observed upper 
limit by ATLAS from the $8$ TeV data corresponding to an integrated 
luminosity of $20.3~{\rm fb^{-1}}$ using the vector boson fusion mode on 
$Br(h \to invi.)$ is $\sim 28\%$ at $95\%$ C.L.~\cite{Aad:2015txa}.  The 
CMS (ATLAS) studies for the high luminosity LHC~\cite{Dawson:2013bba} 
project that one could reach sensitivities for the inivisible branching 
ratio of the Higgs in the range  $17$--$28\%$ ($23$--$32\%$) for 300 
fb$^{-1}$ and $6$--$17$\% ($8$--$16$\%) for 3000 fb$^{-1}$ of integrated 
luminosity for the production of the Higgs in association with a $Z$. More 
recent studies, performed for the future collider 
workshop~\cite{hinv:projected,hinv:projected_2016,Peters:2017olb}, 
possibly reach values as low as $9\% (8\%)$ at $95\%$ C.L. for $3000$ 
fb$^{-1}$ at CMS (ATLAS). It should be noted however, that these smaller 
numbers are usually arrived at by assuming a projected reduction in both 
the systematic and theoretical errors. The more conservative limits  
assuming the same systematic errors as in the current analysis, for CMS 
for example, are around $21 \%$ and $20\%$ for $300$ and $3000$ fb$^{-1}$ 
of integrated luminosity~\cite{CMS-DP-2016-064}, respectively.  Global 
fits to the Higgs coupling data can also probe the invisible branching 
ratio `{\it indirectly'} e.g. ~\cite{Belanger:2013kya}. These limits are 
usually much stronger than those given by the 'direct' searches mentioned 
above  and it is projected that one can reach a sensitivity  of about 5 
$\%$~\cite{Peters:2017olb} at the high luminosity LHC. It should be noted, 
however, that in this way of restricting the '{\it invisible}' branching 
ratio, truly invisible decays are not distinguished from other undetected 
decay modes. Further, these limits are  subject to assumptions on the 
total Higgs decay width, which when modified can lead to different 
results. Hence the limits given by 'direct' searches for an invisibly 
decaying Higgs boson are more model independent and less ambiguous. $e^{+} 
e^{-}$ colliders offer the best possibility of probing such an invisible 
decay mode. The future linear collider ILC, offers the possiblity to probe 
the invisible Higgs branching as low as 0.4$\%$ 
directly~\cite{Asner:2013psa}. The question then naturally arises whether 
such precise measurements will be sufficient to probe light DM models 
(here light refers to  $m_{DM} \lesssim m_h/2$). 

In R-parity conserving Supersymmetry (SUSY),  the lightest stable SUSY 
particle (LSP), typically the neutralino $\chi^0_1$, naturally provides a 
DM candidate. When the neutralino LSP is light enough, $m_{DM} \lesssim 
M_{h}/2$, the Higgs has an invisible decay width into a pair of 
neutralinos. There exist a number of studies addressing the question of 
Higgs decaying invisibly to light neutralinos 
\cite{Griest:1987qv,Djouadi:1996mj,Belanger:2000tg,Belanger:2001am, 
Belanger:2003wb,Calibbi:2011ug,Dreiner:2012ex,Ananthanarayan:2013fga, 
Calibbi:2013poa,Belanger:2013pna,Han:2014nba, 
Belanger:2015vwa,Hamaguchi:2015rxa}.
 Thus there is a direct connection between 
invisible Higgs, DM, and collider signatures in SUSY.
In Higgs portal scenarios, a direct connection between the invisible Higgs 
and direct detection cross-section was established, showing the importance 
of the invisible width for very light DM~\cite{Djouadi:2012zc}. Moreover, 
in simple extensions of the SM, e.g. with an additional scalar singlet, 
the requirement of sufficient annihilation in the early universe to meet 
the relic density constraint means that the coupling of the LSP to the 
Higgs is such that it implies significant invisible width. An exception to 
this requirement is the special case where mass of the DM is just below 
$m_h/2$. Here we wish to revisit the case of neutralino DM  where similar 
arguments apply.

In SUSY, constrained models like the minimal supergravity or the 
constrained minimal SUSY (mSUGRA/cMSSM), the Higgs mass and SUSY searches 
strongly constraint any light neutralino even before imposing DM 
constraints~\cite{Buchmueller:2010ai,Baer:2012uya,Strege:2012bt,Ghosh:2012dh,Roszkowski:2014wqa,Buchmueller:2015uqa,Bechtle:2015nua,Baer:2016ucr,Ellis:2016tjc}. 
In the agnostic, phenomenological Minimally Supersymmetric Standard Model 
(pMSSM), despite the larger number of free parameters, the combination of 
collider, relic abundance and direct detection  results severely restrain 
the possibility of a light neutralino. The argument goes as follows. 
Because of direct limits on charged particles, the light neutralino has to 
be dominantly bino. 
The  annihilation cross-section of the bino is typically not large enough 
to  ensure compatibility with the observed relic abundance unless special 
mechanisms that require some specific mass relations come into play such 
as  annihilation through a Z-boson,  a Higgs or a light pseudo-scalar 
resonance~\cite{Hooper:2002nq,Bottino:2002ry,Belanger:2003wb,Vasquez:2010ru,Feldman:2010ke,Grothaus:2012js,Hamaguchi:2015rxa,Profumo:2016zxo} 
or exchange of light 
sfermions~\cite{Dreiner:2009ic,AlbornozVasquez:2011yq,Arbey:2013aba,Calibbi:2013poa}. 
Thus, generally the scenarios are fine-tuned. 
Recently, a number of studies have considered the status of the pMSSM
 post 7 and/or 8 TeV  runs of the  LHC 
\cite{Belanger:2013pna,Choudhury:2013jpa,Fowlie:2013oua,Cahill-Rowley:2014twa,
Calibbi:2014lga,Bramante:2015una,Ricciardi:2015iwa,Chun:2016cnm,Cao:2015efs,Baer:2016ucr,Badziak:2017the,Beneke:2016ync,Belanger:2000tg,Belanger:2001am} 
and more specifically of the light neutralino.
It was found that  a light neutralino in the pMSSM models is generally 
constrained to masses above $\approx $ 30 GeV 
\cite{Belanger:2013pna,Cahill-Rowley:2014twa,Cao:2015efs} in order to 
avoid over-abundant relic density of the DM although there remains a small 
window at lower masses when the model contains also very light sleptons or 
sbottoms that may escape the LEP 
limits~\cite{Arbey:2013aba,Boehm:2013gst}.
 These light neutralinos can lead to a large branching ratio for the Higgs 
decaying to  invisible particles~\cite{Cahill-Rowley:2014twa} and can be 
further probed in direct detection~\cite{Profumo:2016zxo}, and as well as 
in a collider environment.

Clearly, the precise determination  of the DM relic abundance plays a 
crucial role in constraining light neutralino  DM.  However, the 
constraints are only valid within the framework  of a standard 
cosmological scenario which assumes that  neutralinos have been produced 
thermally and were in thermal equilibrium with SM particles in the early 
universe before they decouple at a   freeze-out  temperature $T_{F}\approx 
m_{DM}/20$. 
Going beyond the simplest assumption of the standard cosmological 
scenario, or more generally not requiring that all the DM be explained by 
the freeze-out mechanism, will clearly open up the possibility for light 
DM in the MSSM. Such scenarios are characterized by a late decaying heavy 
field, for example a SUSY modulus scalar \cite{Moroi:1999zb}. Depending on 
whether a DM candidate in the conventional, thermal freeze-out scenario is 
under-abundant or over-abundant, the non-thermal mechanisms by which one 
attains the observed relic density vary \cite{Gelmini:2006pw}. In SUSY 
with over-abundant DM (for example bino-like DM), the late decay of a SUSY 
modulus scalar field  dilutes the entropy density of the universe. As long 
as the branching fraction of the decaying scalar into DM is not too large,  
the observed relic density can be reproduced for almost all values of 
scalar mass and reheating temperature ($T_{RH}>T_{BBN}$) 
\cite{Gelmini:2006pw,Allahverdi:2012gk}. In the case of thermally under-
abundant DM too, late decay of a scalar can lead to the correct relic 
abundance. 
Thus, both under-abundant or over-abundant DM in the thermal freeze-out 
picture can be brought into agreement with the observed DM abundance. Note 
of course that in the former case there exists the possibility that 
instead of the late decaying scalar it is the existence of multicomponent 
DM that guarantees the correct relic density.

Given the sensitivity of future experiments, both collider and 
astrophysical, it is thus important to re-assess the possibility of 
discovery of  non-thermal light dark matter within the framework of the 
MSSM.  Some studies of non-thermal dark matter in MSSM have been 
conducted~\cite{Gelmini:2006pw,Baer:2014eja,Aparicio:2015sda,Aparicio:2016qqb}. 
The issue of how one can distinguish thermal and non-thermal mechanisms by 
exploiting the complementarity between various experiments has begun to be 
explored eg.~\cite{Roszkowski:2017dou}. Our goal here is  to provide a 
comparative study of the status of light neutralino dark matter in the 
MSSM within both the thermal and non thermal regimes. 
We will consider mainly two types of scenarios : in the first, dubbed 
thermal DM, we assume  the standard cosmological scenario but allow the DM 
to be under-abundant, the underlying assumption is that  another particle 
would form a second DM component or there exists an underlying non-thermal 
mechanism that brings the relic density of an otherwise under-abundant DM 
in agreement with the observed value. In the second scenario, we 
concentrate on the case where thermal production of DM leads to  over-
abundance which clearly calls for a non-standard mechanism. In this case 
we call it a non standard cosmological dark matter (NSDM) scenario. 

 In the framework of the MSSM, we define the parameter space compatible 
 with both classes of DM scenarios. We will show that
 including the possibility of non-thermal DM cosmology opens up the region 
with a light neutralino which  can therefore  contribute  to invisible  
Higgs  decays. After defining the currently allowed parameter space, we 
find the reach  for the  ILC to probe  the remaining parameter space 
either through  Higgs invisible decays or direct production of 
charginos/neutralinos and explore the implications for present and future 
direct detection experiments in both spin dependent and independent 
searches. 
 We further investigate the complementarity of the different collider and 
 direct detection searches to probe the light neutralino scenarios. 
 We shall also address the question of what additional signatures one 
 could primarily focus on at the LHC to achieve this goal, and whether 
 such a signal, in conjunction with 
 the above observations, can decipher the nature of the dark matter 
 particle. We also provide a road-map for the experimental searches for a 
 light neutralino in the MSSM for different possible cosmological 
 histories of DM.

This paper is organized as follows. In Sec.~\ref{sec:par_space_scan}, we 
define the free parameters of the model and the basic collider constraints 
including LEP limits, flavor observables and Higgs physics. 
In Sec.~\ref{sec:collider}, we reinterpret the null results of searches 
for electroweakinos at the LHC. The results of the scan of the parameter 
space for scenarios with a thermal DM that is not over-abundant are 
presented in Sec.~\ref{sec:thermal}, while Sec.~\ref{sec:non_standard} 
includes all non-standard cosmological DM (NSDM) scenarios. 
Sec.~\ref{sec:comp} discusses the potential to probe light neutralinos at 
ILC and in direct detection. In Sec.~\ref{sec:lhc}, we briefly discuss the 
role of high luminosity LHC in probing light neutralino DM. We finally 
conclude in Sec.~\ref{sec:conclusion}.

\section{Model, parameter space and constraints}\label{sec:par_space_scan}

We work within the framework of the MSSM with parameters defined at the 
electroweak scale. Since our main focus is the physics of the Higgs and  
electroweakino sectors, we consider only the nine parameters that capture 
the relevant physics: the gaugino masses $M_1,~M_2$, the higgsino mass, 
$\mu$,  the ratio of the Higgs vacuum expectation values, $\tan\beta$, the 
mass of the third generation squarks, $m_{\tilde{Q}_{3}} $ 
($m_{\tilde{Q}_{3l}},~m_{\tilde{t}_{R}},~ m_{\tilde{b}_{R}}$), the 
trilinear coupling of the stop $A_t$ and  the mass of the gluino $M_3$. 
The first four parameters determine the electroweakino masses and 
couplings while the latter three enter the higher-order corrections to the 
Higgs mass. Note that the mass of the third generation squarks, $ 
m_{\tilde{Q}_{3l}},~m_{\tilde{t}_{R}},~{\rm and}~m_{\tilde{b}_{R}}$, have 
been independently varied between $800~{\rm GeV}$ and $10~{\rm TeV}$.
The two couplings that will be most relevant for our study are those of 
the LSP to the Higgs and to the Z-boson. Both the couplings play a role in 
computing the dark matter observables, while the former also determines 
the invisible width of the Higgs boson. These couplings are defined  
\begin{equation}
\begin{split}
g_{Z\lspone\lspone}=\frac{g}{2 \cos\theta_W} \left( |N_{13}|^2- |N_{14}|^2\right) \\
g_{h\lspone\lspone}=g \left( N_{11}-  \tan\theta_W  N_{12}\right)  \left( 
\sin\alpha  N_{13} +\cos\alpha  N_{14}\right)
\label{eq:coupling}
\end{split}
\end{equation}
where $g$ is the SU(2) coupling, $\alpha$ is the Higgs mixing angle, and 
$N_{1i}$ are elements of the neutralino mixing matrix with $N_{11}$ and 
$N_{12}$ representing the bino and wino components, respectively, while 
$N_{13},N_{14}$ being representatives of the higgsino components.

We explore the reduced nine-dimensional parameter space  of the MSSM with 
a  random  scan within the following ranges:
\begin{eqnarray}
1~{\rm GeV}~<~&& M_{1}~<~100~ {\rm GeV}, \quad 90~{\rm GeV}~<~ M_{2}~<~3~ {\rm TeV},\nonumber\\
 \quad  1~<~&&\tan{\beta}~<~55,\quad 70~{\rm GeV}~<~ \mu~<~3~ {\rm TeV}, \nonumber \\
 \quad 800~{\rm GeV}~<~ m_{\tilde{Q}_{3l}} < 10~{\rm TeV}&&, \quad 
800~{\rm GeV}~<~ m_{\tilde{t}_{R}} < 10~{\rm TeV}, \quad 800~{\rm GeV}~<~ 
m_{\tilde{b}_{R}} < 10~{\rm TeV}, \nonumber \\ 
\quad 2~{\rm TeV}~<~&&M_{3}~<~5~ {\rm TeV}, \quad -10~{\rm TeV}~<~ A_{t}~<~10~ {\rm TeV}
\label{Parameter_space}
\end{eqnarray}
Note that the mass of the gluino is assumed to be large in order to safely 
avoid strong constraints from the LHC. Similarly
the masses of the first and second generation squarks, $m_{\tilde{Q}_{2l},
\tilde{Q}_{1l},\tilde{c}_{L,R},\tilde{s}_{L,R}}$ and the  sleptons masses 
are fixed at 3 TeV,  heavy enough to decouple from the collider 
phenomenology. Moreover $A_{b} = A_{\tau} = 0$ and the pseudoscalar mass 
is taken to be rather heavy, $M_{A} = 1~{\rm TeV}$.  Allowing lower values 
for $M_A$ could impact the dark matter phenomenology, in particular by 
offering a new channel for efficient  light neutralino annihilation 
through a pseudoscalar exchange~\cite{Calibbi:2011ug}. A light second 
Higgs doublet playing a role in neutralino annihilation is however  
strongly constrained from both astrophysical measurements such as FermiLAT 
and LUX~\cite{AlbornozVasquez:2011yq} as well as  from colliders including  
direct searches at the LHC~\cite{Khachatryan:2014wca}, properties of the 
125 GeV Higgs boson~\cite{Bechtle:2014ewa} and searches for the decay 
$B_s\rightarrow \mu^+\mu^-$~\cite{Aaij:2012nna}.   
 We generate $\approx 10^7$ parameter space points and implement the 
 relevant constraints in order to obtain the allowed parameter space.  We 
 only consider points for which the decay of the Higgs into a pair of 
 neutralinos is kinematically allowed. 

The parameter space is initially constrained by imposing the limits on 
mass of the light CP-even neutral MSSM Higgs boson ($h$). Note here that 
we impose that the  lightest Higgs $h$ behaves like the SM Higgs boson 
with a mass of 125 GeV, with the couplings satisfying the constraints 
derived from the LHC.  
We also constrain the parameter space by imposing  low energy flavor 
physics constraints, the limit on the invisible decay width of Z boson, 
LEP limits on electroweakinos, 
and Higgs signal strengths limits  derived from the LHC Run-I data, 
through a combined analysis by ATLAS and CMS collaborations. These 
constraints are discussed in more detail below, while other constraints 
from direct electroweakino searches at the LHC are detailed in the next 
section. 

\begin{itemize}
\item \textbf{Light Higgs mass} : ATLAS and CMS have performed a combined 
measurement of the Higgs mass ($M_{h}$) and its value has been determined 
to be in the range $124.4-125.8$ GeV \cite{Aad:2015zhl} at $3\sigma$.
Taking into account the  theoretical uncertainties associated with   the  
computation of the  Higgs mass, we choose a conservative
 approach and impose that the light Higgs mass lies within the range, 
 $122~{\rm GeV} < M_{h} <  128~{\rm  GeV}$.  
The particle spectrum  is generated using SUSPECT (version 2.43) 
\cite{Djouadi:2002ze} and includes dominant 2-loop corrections to the 
Higgs mass in the $\overline{DR}$ scheme.  

\item \textbf{Flavor Physics Observables} : Flavor physics observables are 
among the most sensitive probes of new physics effects and have been 
extensively used to constrain BSM physics.
Since the mass of the first two generations of squarks have been fixed at 
a high value of 3 TeV, their contributions to the rare-decay processes 
decouple. Thus the main contribution to $B_{s} \to \mu^{+} \mu^{-}$  comes 
from penguin diagrams with the incoming b and s quarks coupled to a 
chargino and an up-type squark.
These branching fractions were obtained using micrOMEGAs (version 4.2.3) 
\cite{Belanger:2004yn,Belanger:2001fz,Belanger:2014vza}
We adopt a moderate approach here, and impose the constraints on the 
branching fraction of $B_{d} \to X_{s} \gamma $ and $B_{s} \to \mu^{+} 
\mu^{-}$, allowing $2\sigma$ uncertainty with respect to the currently 
measured best-fit values : $B_{d} \to X_{s} \gamma = 
(3.32\pm0.15)\times10^{-4}$ \cite{Amhis:2016xyh} and  $B_{s} \to \mu^{+} 
\mu^{-} =  2.8 ^ {+0.7}_{-0.6}  \times 10^{-9} $ \cite{CMS:2014xfa}.

\item \textbf{LEP limits} :  Upper limits have been derived from LEP data 
on the associated neutralino production cross-section ($\sigma_{\lspone 
\lsptwo}$) times the branching fraction of $\lsptwo \to q \tilde{q} 
\lspone$ \cite{Abbiendi:2003sc}.
An upper limit has been obtained on $\sigma_{\lspone \lsptwo}~<~0.1~{\rm 
pb}$ \cite{Abbiendi:2003sc} at $95\%$ C.L., for $(|M_{\lsptwo} - 
M_{\lspone}| > 5~{\rm GeV})$. We implement this constraint in our 
analysis, and the values of the relevant variables have been obtained 
using micrOMEGAs (version 4.2.3) 
\cite{Belanger:2004yn,Belanger:2001fz,Belanger:2014vza}. We also impose an 
upper limit on the invisible decay width of Z boson, 
$\Gamma_{Z}^{inv}<2~{\rm MeV}$ \cite{ALEPH:2005ab}. Here "invisible" 
refers to the non-SM invisible decay modes only. In addition, we also 
impose a lower limit on chargino mass, $M_{\chonepm} < 103~{\rm GeV}$ 
\cite{Abbiendi:2003sc}. This constraint implies that the lightest 
neutralino with a mass below $M_{h}/2$  be dominantly  bino.

\item \textbf{Higgs boson width} : We also impose an upper limit on the 
total decay width of the Higgs boson, $\Gamma_{h} < 22 ~{\rm 
MeV}$\cite{Khachatryan:2014iha}, derived by CMS at $95\%$ C.L., using LHC 
data collected at $\sqrt{s}=7~{\rm TeV}$ and $8~{\rm TeV}$ corresponding 
to an integrated luminosity of $5.1~{\rm fb^{-1}}$ and $19.7~{\rm 
fb^{-1}}$, respectively.  

\item \textbf{Higgs signal strength constraints:} The Higgs boson is 
dominantly produced at LHC through gluon fusion ($ggF$). The other 
productions modes are vector boson fusion ($VBF$), associated production 
with a vector boson ($Vh$, where, $V = W,Z$), associated production with a 
pair of top quarks ($t\bar{t}h$). Both, ATLAS and CMS collaborations have 
analysed various final states from Higgs boson decay, like $h \to \gamma 
\gamma,~W^{+}W^{-},~ZZ,~b\bar{b},~\tau^{+}\tau^{-}$, and have presented 
their results in the form of signal strength variables $\mu^{i}_{f}$, 
where $\mu^{i}_{f}$ is defined as the ratio of Higgs production cross-
section in the $i-th$ production mode times the branching fraction of 
Higgs boson in the $f$ final state for the model under consideration, with 
respect to the same quantity in SM. It is expressed as:\begin{eqnarray}
\mu^{i}_{f} = \frac{\sigma_{i} \times Br(h \to f)}{{(\sigma_{i} \times 
Br(h \to f))}_{SM}}
\end{eqnarray} 
Here, $i$ corresponds to the production modes of Higgs boson, 
$i=ggF,~VBF,~Vh,~t\bar{t}h$, and f corresponds to the decay channel of 
Higgs boson, $f = \gamma 
\gamma,~W^{+}W^{-},~ZZ,~b\bar{b},~\tau^{+}\tau^{-}$.

ATLAS and CMS have performed a combined analysis of LHC Run-I data and 
have presented their results in the form of two-dimensional correlation 
contours in the $\mu_{ggF+Vh} - \mu_{VBF+t\bar{t}h}$ plane at $68\%$ and 
$95\%$ C.L., shown in Fig.~28 of \cite{ATLAS-CONF-2015-044}. In this work, 
we consider only those parameter space points which lie inside the $95\%$ 
C.L. correlation contours for the $\gamma 
\gamma,~W^{+}W^{-},~ZZ,~b\bar{b},~\tau^{+}\tau^{-}$ final states. Note 
that the currently available $13~{\rm TeV}$ data is comparable to the 
$8~{\rm TeV}$ data~\cite{Barman:2016jov}. 

\item \textbf{ Constraints from the invisible Higgs decays:} We also take 
into account direct search limits on the invisible 
decay of the Higgs boson. 
Direct search limits for an invisible decay mode of the Higgs in the Wh, 
Zh and the VBF channel performed by the ATLAS collaboration using the full 
7 and 8~TeV data, sets the upper limit 
to be 0.25 \cite{Aad:2015pla}. The CMS collaboration similarly, by using 
the full 7, 8 TeV, as well as 2.3 fb$^{-1}$ data from the 13 TeV run sets 
an upper limit of 0.24 from direct searches on the invisible branching 
ratio of the Higgs \cite{CMS-PAS-HIG-16-016}.  Note that in all of the 
above, standard model production cross-section and BR is assumed.
 In practice,  the scenarios satisfying  the Higgs signal strength 
 constraints also automatically respect the direct limit constraints. 

\end{itemize}

\section{Constraints from electroweakino searches at the LHC}\label{sec:collider}

Having documented the constraints on the electroweakino parameter space 
from LEP and other indirect sources in the previous section, 
we turn our attention to direct collider searches, focussing on the region 
of parameter space of interest in this study. 
Within the scope of this work, the LSP mass $M_{\lspone} \leq M_{h}/2$, 
and hence, the LEP constraints impose that the light LSP be dominantly 
bino. On the other hand, the relic density constraint
requires that the light LSP also possess a higgsino or wino 
component~\cite{Hooper:2002nq,Bottino:2002ry,Belanger:2003wb,Vasquez:2010ru,Feldman:2010ke,Grothaus:2012js,Profumo:2016zxo}.
Since we will consider both, the case of the neutralino being a standard 
thermal relic, and the case of alternative cosmologies, for collider 
studies 
we remain agnostic to the finely tuned relic density viable regions.  
Within the framework of a model with heavy sfermions, 
the most relevant parameters are  $\mu,M_{1},M_{2}$ and $\tan{\beta}$ 
which determine the electroweakino couplings to Higgs and gauge bosons
and therefore influence both the production cross-section, in particular 
for Drell-Yan processes, and the dominant decay modes.
 As an example, the decay of $\chi_{i}^{0}\to \chi_{1}^{0}h$  is strongest 
 when one of the neutralino is gaugino-like and the 
other one higgsino-like, while the decay $\chi_{i}^{0}\to \chi_{1}^{0}Z$ 
is determined only by the higgsino components of the neutralinos, see 
Eq.~\ref{eq:coupling}.

The processes of interest for this study are 

\begin{equation}\label{dilepton}
p p \to \chi_{i}^{+}\chi_{j}^{-} \to l^{+}l^{-} + \met
\end{equation}

\noindent
and
\begin{equation}\label{trilepton}
p p \to \chi_{i}^{+}\chi_{j}^{0} \to 3\ell + \met
\end{equation}

\noindent
where, $\met$ represents the missing transverse energy arising from the 
$\lspone$ and neutrinos.

The above processes can occur via  a) direct decays  
$\chi^{\pm}_{i}\to\chi_{1}^{0}W^{\pm} $ and 
$\chi_{i}^{0}\to\chi_{1}^{0}Z/h$  or b) cascade decays 
of higher chargino and neutralino states.
 Depending on the mass gap between the $\chi_{i}^{\pm}\chi_{i}^{0}$  and 
 $\chi_{1}^{0}$,   the W/Z/h bosons can be on or off-shell. 

 Since our primary motivation in this section is to assess the LHC 
 constraints on  light neutralinos, 
 we choose 3 discrete values for $M_1$ ($M_1=5,40,60{\rm~ GeV}$) in order 
 to cover the range of relevant LSP masses.
 Note that the mass of the lightest neutralino is roughly determined by 
 the value of $M_1$, with corrections due to the mixing of other 
 electroweakino mass parameters, which can reach a few GeV's especially 
 when $\mu$ is also small.
 We then perform  a scan in the $\mu-M_{2}$ plane and determine the 
 constraints from the dilepton/trilepton + missing transverse energy (MET) 
 searches performed at the LHC in the 8 TeV run.  
To this end we use the recasted LHC 8 TeV results in publicly available 
analyses databases in the framework of MadAnalysis5 \cite{Dumont:2014tja} 
and Checkmate \cite{Dercks:2016npn}. 
We do not include the currently available 13~TeV results 
\cite{Aad:2016tuk} since a faithful recast of the corresponding 
electroweakino searches are not yet publicly available, however, we will 
briefly discuss below about the prospects of improving the collider limits 
with higher luminosity.

The first search of interest is the ATLAS search for direct production of 
chargino and neutralino in the 3 lepton + MET channel as documented in 
\cite{Aad:2014nua}. The search is optimized on several scenarios, 
including light slepton or light staus in the intermediate state, however 
the WZ/Wh mediated scenario is the only one of relevance here.  
The basic selection criteria for the signal region requires one pair  of 
same flavor opposite sign leptons (SFOS) among the three selected leptons 
with a transverse momentum ($p_{T}$) threshold of $25$ GeV. The WZ 
mediated signal region has 20 disjointed bins based on the MET, the 
invariant mass of the same flavor opposite sign leptons and 
the transverse mass between the 3rd lepton and MET. The other signal 
regions in this analysis follow a similar pattern. This analysis has been 
validated 
in the Checkmate framework, the details of which can be found in 
\cite{Dercks:2016npn}. 

The second search of interest is the ATLAS search for direct production of 
charginos  and neutralinos in the dilepton + MET final state 
\cite{Aad:2014vma}. The opposite sign  
 dilepton + MET signature arises from chargino pair production followed by 
 the decay to the LSP and  leptons, via intermediate state W bosons. 
 Signal regions are divided into two criteria, first requiring a veto on 
 the 
Z boson and the second without. This analysis was recasted and validated 
in the public analysis database (PAD) framework of MadAnalysis5. The 
details of the recasted analysis, the validation procedure and the recast 
code can be found in \cite{maweakino}.  As before, the validation was 
found to be reliable and hence we use this 
analysis to constrain the parameter space. 

Finally we also consider monojet + MET searches. It is well known that 
these searches work well  for compressed  decay topologies. However, it 
has also been noted that it could probe light DM for certain classes of 
models \cite{Aaboud:2016tnv}. Thus, we assess whether the constraints
in regions where the decay of charginos and neutralinos via off shell W/Z 
boson lead to soft leptons and jets could be improved by mono-jet 
searches. 
For this we use the publicly available recasts of  the ATLAS 8 and 13 TeV 
monojet  + MET searches  \cite{Aaboud:2016tnv,Aad:2014nra} available 
in the PAD  framework. The two ATLAS monojet + MET searches are similar in 
terms of the nature of selection cuts and implementation, the difference 
being the strength of the applied cuts when going from 8 to 13~TeV. The 
analysis relies on the emission of one hard jet at the initial 
state, which recoils against the MET. The ATLAS analysis for both 8 and 13 
TeV is divided into signal regions 
of increasing leading jet transverse momentum and MET. The details of the 
validation is documented and the implemented
 code is available in \cite{mono8,mono13}.

We re-interpret the above searches by generating 100,000 events for each 
point in the $\mu-M_{2}$, plane for three discrete values of $M_{1}$, with 
$\tan\beta$ fixed to 10, corresponding to the processes described in 
Eq.~\ref{dilepton} and \ref{trilepton} using MadGraph5 
\cite{Alwall:2014hca}. The rest of the spectrum, (including the sleptons) 
are decoupled from this set. The events are then passed to PYTHIA6 
\cite{Sjostrand:2006za} for showering and hadronization. Jets are 
reconstructed using FASTJET \cite{Cacciari:2011ma}, with the 
reconstruction parameters chosen to satisfy the requirements of each of 
the above analysis. Detector simulation is performed using Delphes 
\cite{deFavereau:2013fsa} with the detector parameters obtained from the 
validated cards for each of the above analysis. Cross-sections for each 
point are calculated using Prospino~\cite{Beenakker:1996ed}. Note that the 
cross-section varies mildly with $\tan\beta$  and that this does not 
impact the collider bounds significantly.
For each of the above recasts, the exclusion curves are obtained by built-
in confidence level calculators following the CL$_{s}$ prescription 
\cite{Read:2002hq}. In MadAnalysis5, for example,
the module $exclusion-CLs.py$  determines, given the number of signal, 
expected and observed background events, together with the background 
uncertainty (the latter three directly
taken from the experimental publications), the most sensitive signal 
region (SR) of the analysis and the exclusion C.L. using the CLs 
prescription for the most sensitive SR.

\begin{figure}[ht]
\begin{center}
\includegraphics[scale=0.17]{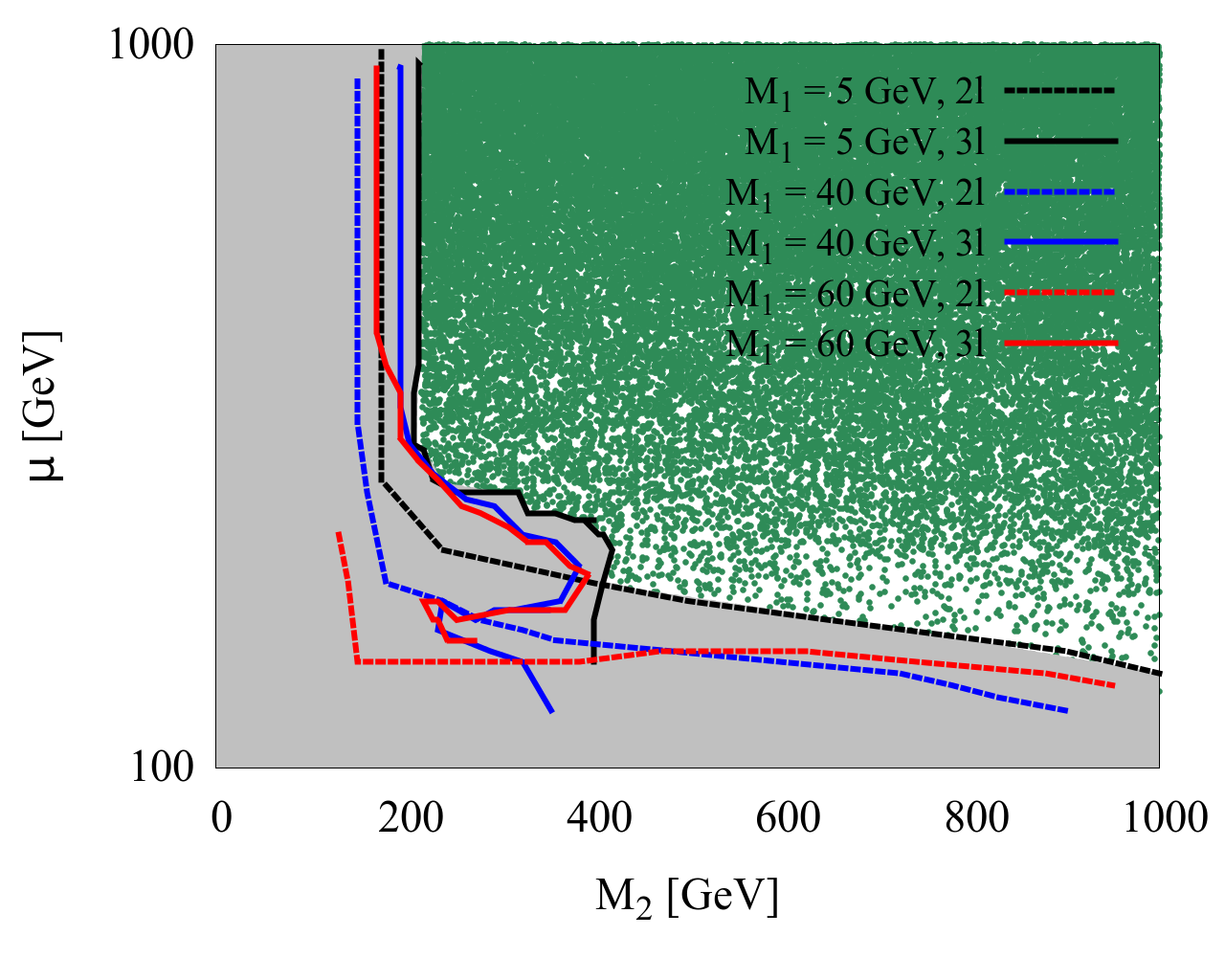} 
\caption{95\% C.L. contours in the $\mu-M_2$ plane  from dilepton (dashed) 
and trilepton (solid) searches  at LHC-8~TeV for $M_1=5,40,60~{\rm GeV}$. 
Here $\tan\beta=10$. Green points correspond to  the allowed points of the 
scan after imposing all constraints in Sec.~\ref{sec:par_space_scan}. Only 
the region $\mu<1$~TeV is displayed, as the contours are independent of 
$\mu$ for low values of $M_2$.  }
\label{LHCcont}
\end{center}
\end{figure}

In Fig.~\ref{LHCcont}, we present  the  95\% C.L. contours  in the $\mu-
M_{2}$ plane for three values of  $M_{1}$. The green shaded region are 
allowed points subject to the constraints on the light Higgs mass, the 
flavor physics constraints, and the LEP limit on the light chargino mass, 
along with the Higgs signal strength correlations from the combined 
CMS+ATLAS analysis.  The solid lines correspond to the constraints from 
the  trilepton search, while the dotted lines correspond to the dilepton 
search. We observe that the trilepton search is more constraining than the 
dilepton search except in the region of large $M_2$. These observations 
are consistent with the results obtained in \cite{Arina:2016rbb}. There 
are various channels that contribute to the exclusions curves in this 
figure. 

For the region at large $M_2$ excluded by the dilepton search, the main 
contributing processes are $p p \to \chi_{2,3}^{0}\chi_{1}^{0}$  with the 
heavier neutralinos decaying into $Z\chi_{1}^{0}$. 
The production cross-sections decrease rapidly as $\mu$ increases and the 
LSP becomes pure bino, thus setting the exclusion. The dependence on $M_1$ 
is basically set by the detection efficiency as the BR of 
charginos/neutralinos into the LSP and a gauge or Higgs boson is nearly 
100\% in all cases.
As $M_2$ decreases,  the process  $p p \to \chi_{1}^{+}\chi_{1}^{-}$ 
contributes significantly to the exclusion, until  for small values of 
$M_2$ when the chargino is dominantly wino it becomes the only relevant 
process. In this region however the best exclusions are set from the 
trilepton search. 
 For the trilepton search,  the majority of the exclusion originates from 
 $p p \to \chi_{1}^{+}\chi_{2/3}^{0}$. At large values of $\mu$, 
 $\chi_{2}^{0}$ is mostly wino and
 $p p \to \chi_{1}^{+}\chi_{2}^{0}$ is the dominant channel, the 
 production cross-section is determined by the value of $M_{2}$ and the 
 exclusion contours are basically independent of $\mu$. For this reason,  
 in Fig.~\ref{LHCcont}, we show only the region up to $\mu=1$~TeV. 
 Note that  in this region, the exclusion is more stringent for lower 
 values of $M_1$.  To a large extent this is due to more available missing 
 energy, thus enhancing the efficiency of the search.
 For lower values of $\mu$ the production cross-sections for both 
 $\chi_{1}^{+}\chi_{2}^{0}$ and $\chi_{1}^{+}\chi_{3}^{0}$ increases thus 
 extending the reach, until some value of $M_2$ where the search loose 
 sensitivity partly because of the low cross-sections involved. 
 Note that these limits are derived using the best expected signal region, 
 however the best signal region jumps in the region of parameter space 
 near the  curve excluded by the trilepton search,  hence the contour is 
 uneven.  Finally  the mono-jet  searches (both from 8 and 13~TeV) do not 
 constrain the parameter space. The reason for this is twofold. Since the 
 dark matter is light, the jet recoiling against this light object is not 
 hard enough, and therefore the acceptance $\times$ efficiency is not large. 
As the production cross-section is low, this inefficiency in acceptance is 
not compensated. We would like to mention here that we have implemented 
the exclusion limits derived at $95\%$ C.L. from the dilepton and 
trilepton searches from LHC 8 TeV data, which are shown in 
Fig.~\ref{LHCcont}, on our parameter space of interest.

 We performed a study based on the cuts designed to probe electroweakinos 
 at the high luminosity LHC \cite{ATL-PHYS-PUB-2014-010}, for low to 
 intermediate values of $\mu$ and $M_{2}$ (200-400 GeV), and observed that 
 the selection cuts for the  search were not optimal to probe this region 
 of the parameter space. Thus, a detailed  study for the high luminosity 
 LHC run is required and will be the subject of a follow up to this work. 
 We emphasize that all the region with charginos lighter than 500 GeV 
 (roughly $\mu,M_2 < 500~ {\rm GeV}$) can be easily probed at a TeV scale 
 ILC.

\section{The neutralino as a Thermal Relic}\label{sec:thermal}

In this section, we consider the case where the neutralino is a thermal 
relic and discuss the impact of DM observables including the relic density 
and the elastic scattering  of DM with nucleons. Within the standard 
cosmological model, the neutralino relic density is computed using 
micrOMEGAs (version 4.2.3) 
\cite{Belanger:2004yn,Belanger:2001fz,Belanger:2014vza} and compared with 
the very precise measurement done by the PLANCK 
collaboration~\cite{Ade:2015xua}, $\Omega_{DM}^{obs}h^{2} = 0.1184 \pm 
0.0012$ at 68$\%$ C.L. Assuming $3\sigma$ interval, we obtain a window of 
$\Omega_{DM}^{obs}h^{2} = 0.1148 - 0.1220$, and adopting a conservative 
approach, we impose an  upper limit, $\Omega_{DM} h^2 \leq 0.1220 $. Here 
we assume implicitly that either there is another DM component when the 
observed value is not saturated, or, the observed value is attained 
through non-thermal mechanisms.
We also impose the constraints from LEP, flavor physics and Higgs physics, 
listed in Sec.~\ref{sec:par_space_scan} and consider only the region where 
the Higgs decay into neutralinos is kinematically accessible.

\begin{figure}[ht]
\begin{center}
\includegraphics[height=2.42in,width=3.0in]{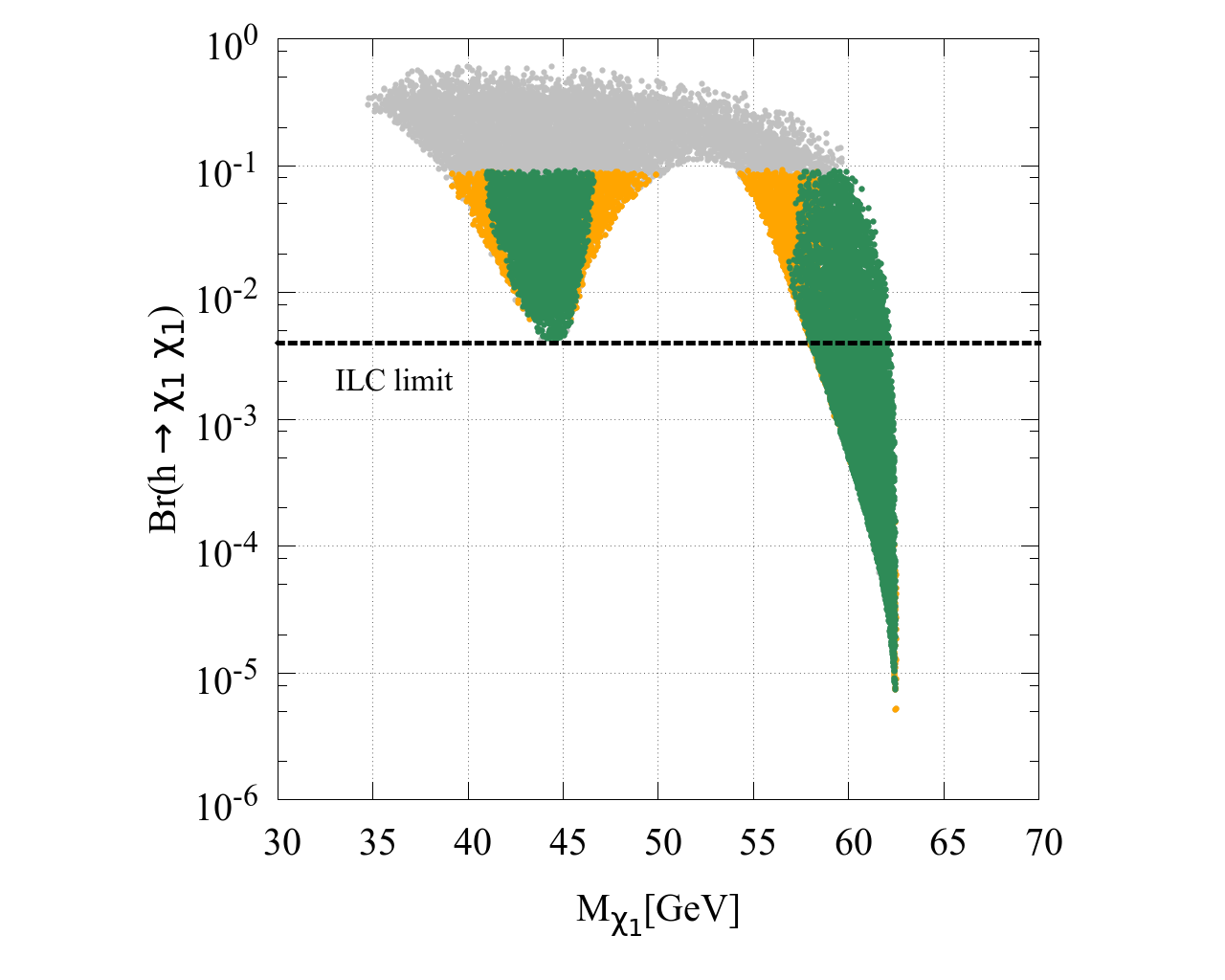}
\includegraphics[height=2.42in,width=3.0in]{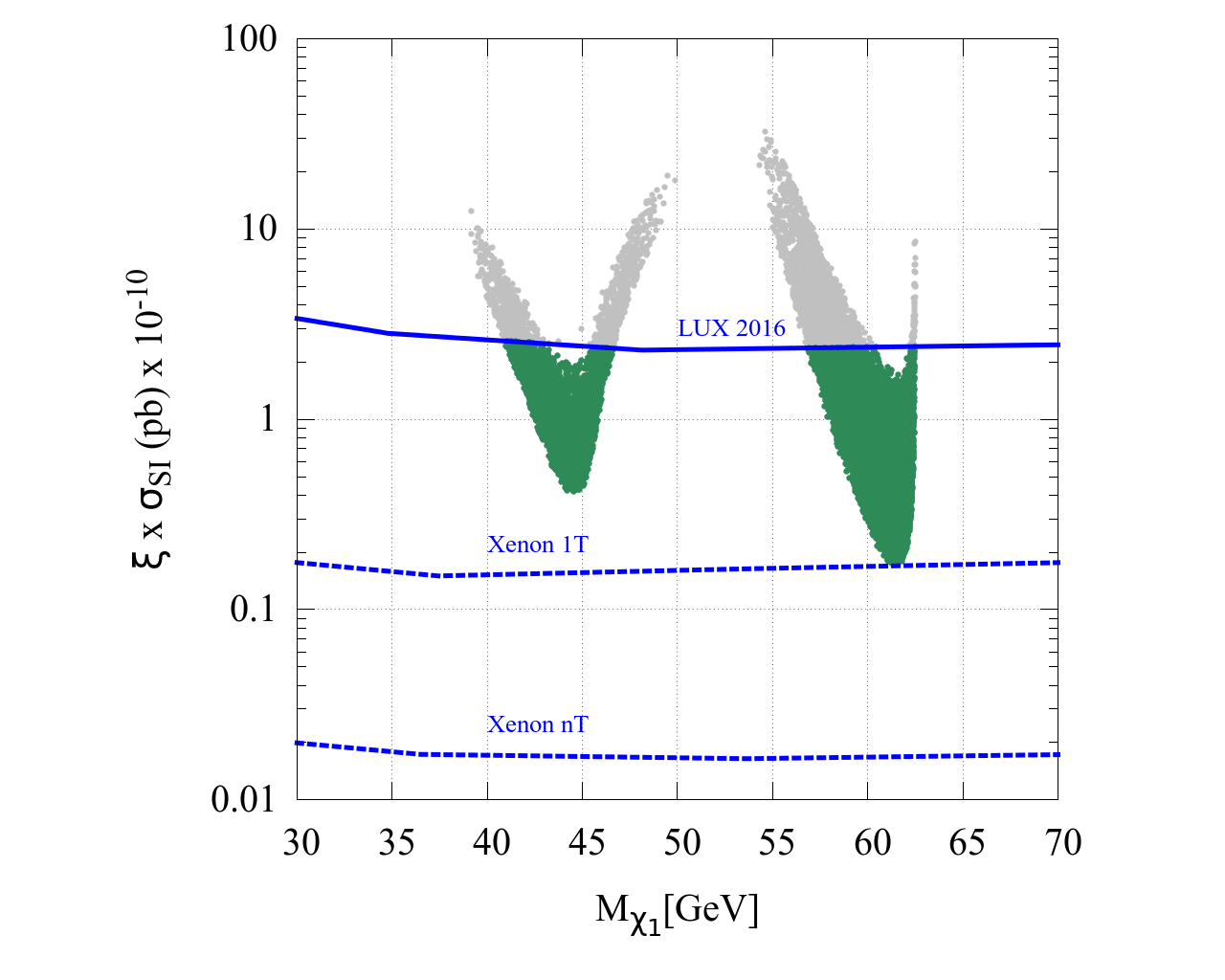}
\caption{ \textbf{a)} The Higgs to invisible branching $Br(h \to \lspone 
\lspone)$ vs. the  LSP mass $M_{\lspone}$. The grey (coloured) points 
distinguish the points allowed before (after) the Higgs signal strength 
constraints. Yellow (green) points are excluded (allowed) by the current 
limits on SI WIMP-nucleon cross-section from 
LUX-2016~\cite{Szydagis:2016few}. The black-dashed line represents the ILC  
reach, $Br(h \to \lspone \lspone)~> 0.4\%$~\cite{Asner:2013psa}.
 \textbf{b.)}  SI WIMP-nucleon cross-section vs $M_{\lspone}$ for all 
 points allowed by collider and relic density constraints. The  blue-solid 
 line show the current limit from LUX-2016~\cite{Szydagis:2016few} and the 
 blue-dashed lines shows the projected reach for  
 Xenon-1T~\cite{Aprile:2015uzo} and  Xenon-nT~\cite{Aprile:2015uzo}.}
\label{Thermal:Br_Inv}
\end{center}
\end{figure}

Imposing the DM relic density bound sets a lower limit on the LSP mass, 
$M_{\lspone} \gtrsim 34~{\rm GeV} $. As mentioned previously, the LEP 
limits on charged particles entail that the neutralino DM,  $\lspone$, be 
bino-dominated with mixtures from higgsino as well as wino. Within our 
framework, the only mechanisms for achieving efficient DM annihilation are 
exchange of a Z or a Higgs boson. As expected, we observe that allowed 
points are restricted to the funnel regions with the LSP mass near $m_Z/2$  
or $m_h/2$.  After taking into account the constraints from the Higgs 
signal strengths which effectively restrict the coupling of the LSP to the 
Higgs, in fact reducing the higgsino component of the LSP thus  also its 
coupling  to the Z, the mass of the LSP is forced to lie even closer to 
either resonance. The impact of the Higgs coupling constraints is 
displayed in Fig.~\ref{Thermal:Br_Inv}(a) which shows the branching 
fraction of the CP-even light Higgs boson (h) to a pair of $\lspone$ 
($Br(h \to \lspone \lspone)$) as function of the LSP mass, $M_{\lspone}$. 
The grey points only satisfy the Higgs mass constraint, the flavor physics 
constraints and LEP limits mentioned in Sec.~\ref{sec:par_space_scan}, 
whereas the coloured points also satisfy the constraints from Higgs signal 
strength measurements. We observe that after applying the latter,
the Higgs to invisible branching fraction is restricted to $\lesssim 
10\%$. A high invisible Higgs branching fraction severely affects its 
branching to the  SM decay modes, resulting in the signal strength values 
receiving a strong shift from their SM values and as a result, falling 
outside the 95$\%$ C.L. Higgs signal strength correlation contours 
discussed in Sec.~\ref{sec:par_space_scan}. We show the direct ILC reach 
in the Higgs to invisible mode ($Br(h \to invi.) > 
0.4\%$~\cite{Asner:2013psa}) through a black dashed line in 
Fig.~\ref{Thermal:Br_Inv}(a). It can be observed from 
Fig.~\ref{Thermal:Br_Inv}(a) that the ILC will be able to probe the entire 
Z funnel region  through the Higgs to invisible branching. 
However, in the Higgs resonance region, the Higgs to invisible branching 
fraction attains value as low as $\approx 10^{-5}$, thus  a significant 
fraction of points  will evade detection by ILC (through the Higgs to 
invisible decay mode).
The reason for the small invisible width is on one hand, the small  LSP-
Higgs coupling  and on the other hand, phase space suppression.  
Finally, note that the points  for which the relic density falls precisely 
within the observed range lie at the lower edge of the colored region in 
Fig.~\ref{Thermal:Br_Inv}(a).

In  recent years, DM direct detection experiments which exploit spin-
independent (SI) or spin-dependent(SD) WIMP-nucleon elastic scattering 
have provided very sensitive probes of DM. 
Since the SI WIMP-nucleon experiments are more sensitive than their SD 
counterpart, in our framework, we display only the predictions and the  
limits from SI WIMP-nucleon interaction in  Fig.~\ref{Thermal:Br_Inv}(b).  
Here, the grey and green colored parameter space points satisfy all the 
constraints mentioned in Sec.~\ref{sec:par_space_scan}, however the grey 
points are excluded by the current LUX limits~\cite{Szydagis:2016few}. 
Typically the excluded points are those with a larger higgsino component 
as the elastic scattering cross-section is dominated by Higgs exchange, 
hence depends directly on the LSP-Higgs coupling. Note that the LUX-2016 
limit permits to exclude many points for which the invisible Higgs width 
is small (even only 1\%) and is therefore more stringent than current 
Higgs precision measurements, see Fig.~\ref{Thermal:Br_Inv}(a). Moreover, 
the upcoming Xenon-1T experiment~\cite{Aprile:2015uzo} will be able to 
probe the entire allowed parameter space, as is evident from 
Fig.~\ref{Thermal:Br_Inv}(b). The exclusion limits are obtained assuming a 
local density of DM $\rho=0.3~ \rm{GeV/cm}^3$, large  uncertainties in its 
determination  
can introduce a large shift in the limit extracted~\cite{Benito:2016kyp}, 
the conventional value used lead  to a somewhat conservative limit.

Note that we have rescaled the SI WIMP-nucleon cross-section ($\sigma_{SI}$) with 
$\xi$, defined as the ratio of predicted DM relic density $\Omega h^2$ to 
the observed one $\Omega_{DM} h^2$ ($0.122$, allowing $3\sigma$ interval 
around the best-fit value derived by PLANCK 
collaboration~\cite{Ade:2015xua}).
\begin{eqnarray}
\xi = \frac{\Omega}{\Omega_{DM}} = \frac{\Omega}{0.122}
\end{eqnarray}

\begin{figure}[ht]
\begin{center}
\includegraphics[height=2.42in,width=3.0in]{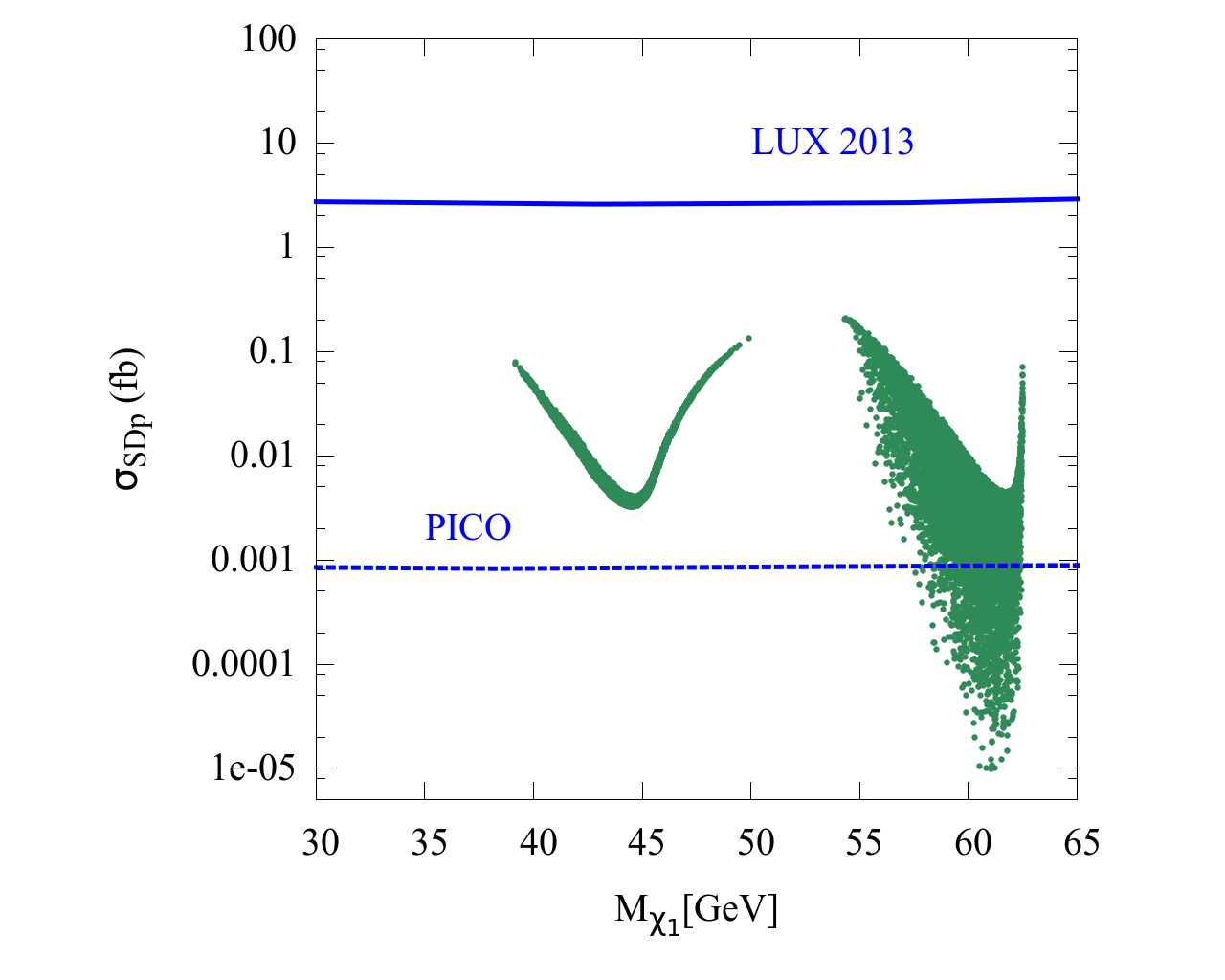}
\includegraphics[height=2.42in,width=3.0in]{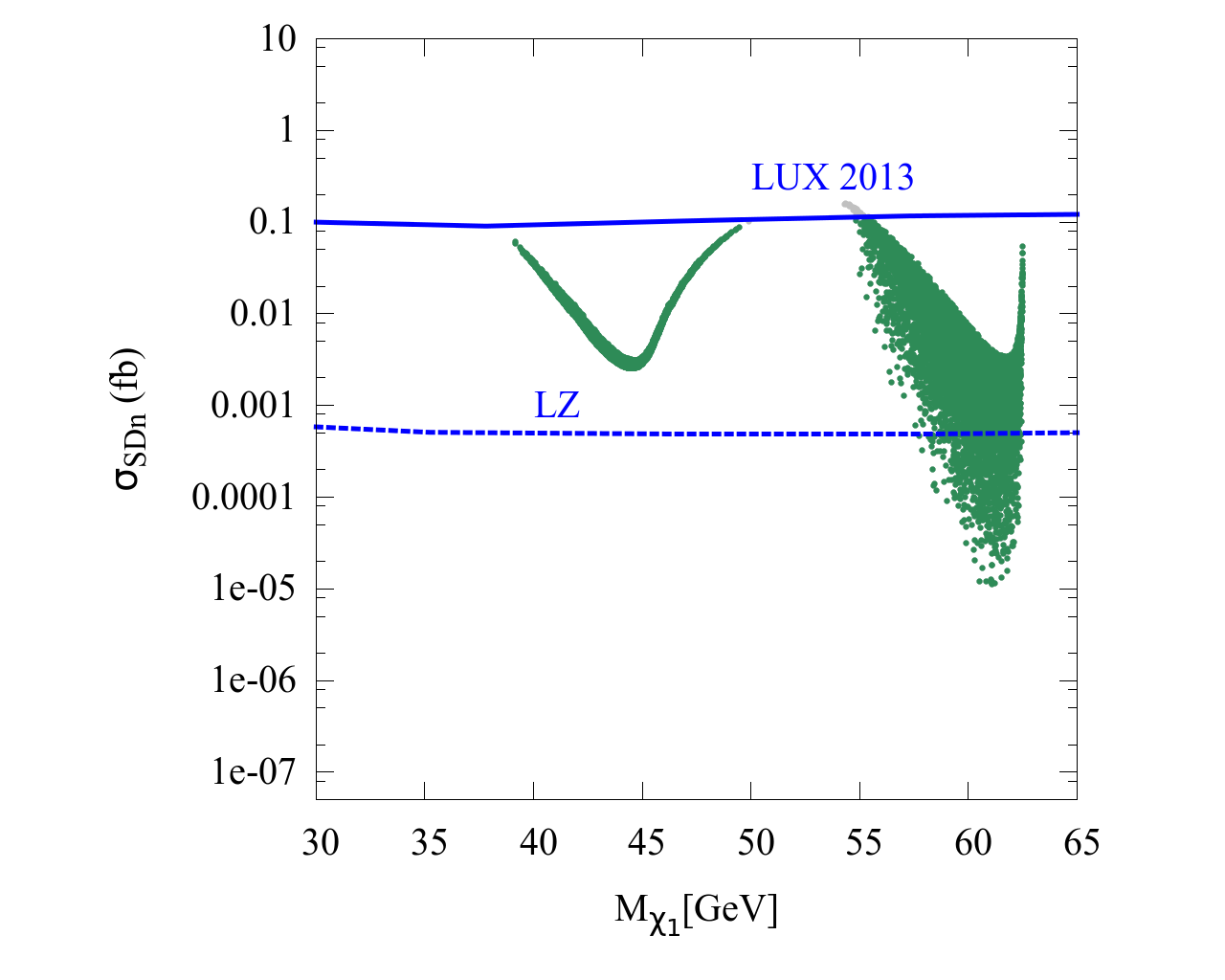}
\caption{ \textbf{a)} SD WIMP-proton cross-section vs $M_{\lspone}$ for 
all points allowed by collider and relic density constraints. The  blue-
solid line show the current limit from LUX-2013~\cite{Akerib:2016lao} and 
the blue-dashed line shows the projected reach for  
PICO-250~\cite{Cushman:2013zza}.
 \textbf{b.)}  SD WIMP-neutron cross-section vs $M_{\lspone}$ for all 
 points allowed by collider and relic density constraints. The  blue-solid 
 line show the current limit from LUX-2013~\cite{Akerib:2016lao} and the 
 blue-dashed line shows the projected reach for LZ~\cite{Akerib:2016lao}.}
\label{Thermal:mu_sd_scale}
\end{center}
\end{figure}

The predictions for the rescaled spin dependent cross-section on protons 
($\sigma_{SD}^{proton}$)  and on neutrons  ($\sigma_{SD}^{neutron}$) are 
shown  in Fig.~\ref{Thermal:mu_sd_scale} together with the current limits 
from LUX~\cite{Akerib:2016lao}  and the future projections from 
PICO-250~\cite{Cushman:2013zza}  for $\sigma_{SD}^{proton}$ and from 
LZ~\cite{Akerib:2016lao} for
$\sigma_{SD}^{neutron}$. Note that the cross-sections on protons and 
neutrons are very similar. PICO-250 and LZ will be able to probe the 
entire Z-resonance region, while part of the Higgs resonance region 
remains out of reach of planned detectors.

\begin{figure}[ht]
\begin{center}
 \includegraphics[height=2.42in,width=3.0in]{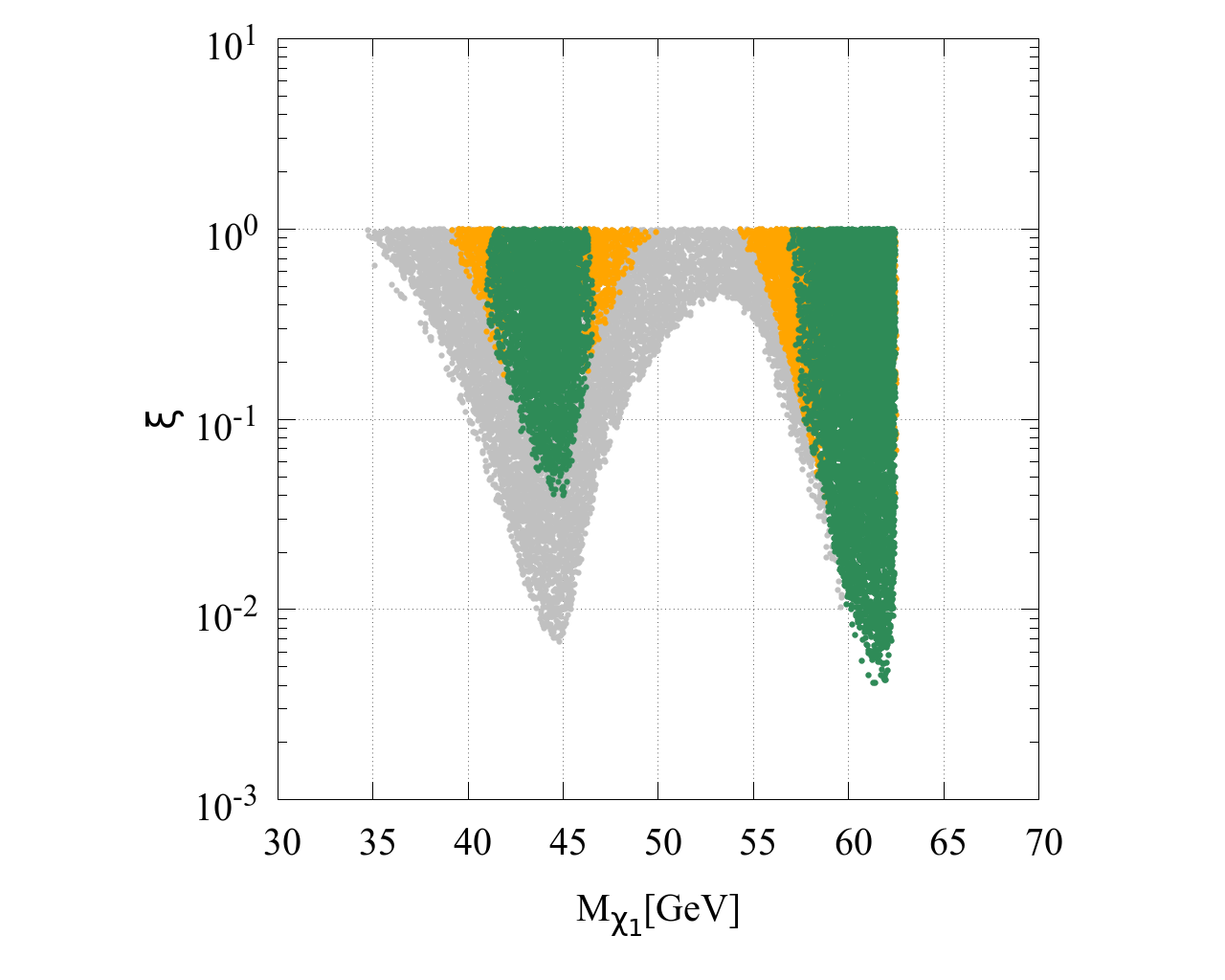}
\includegraphics[height=2.42in,width=3.0in]{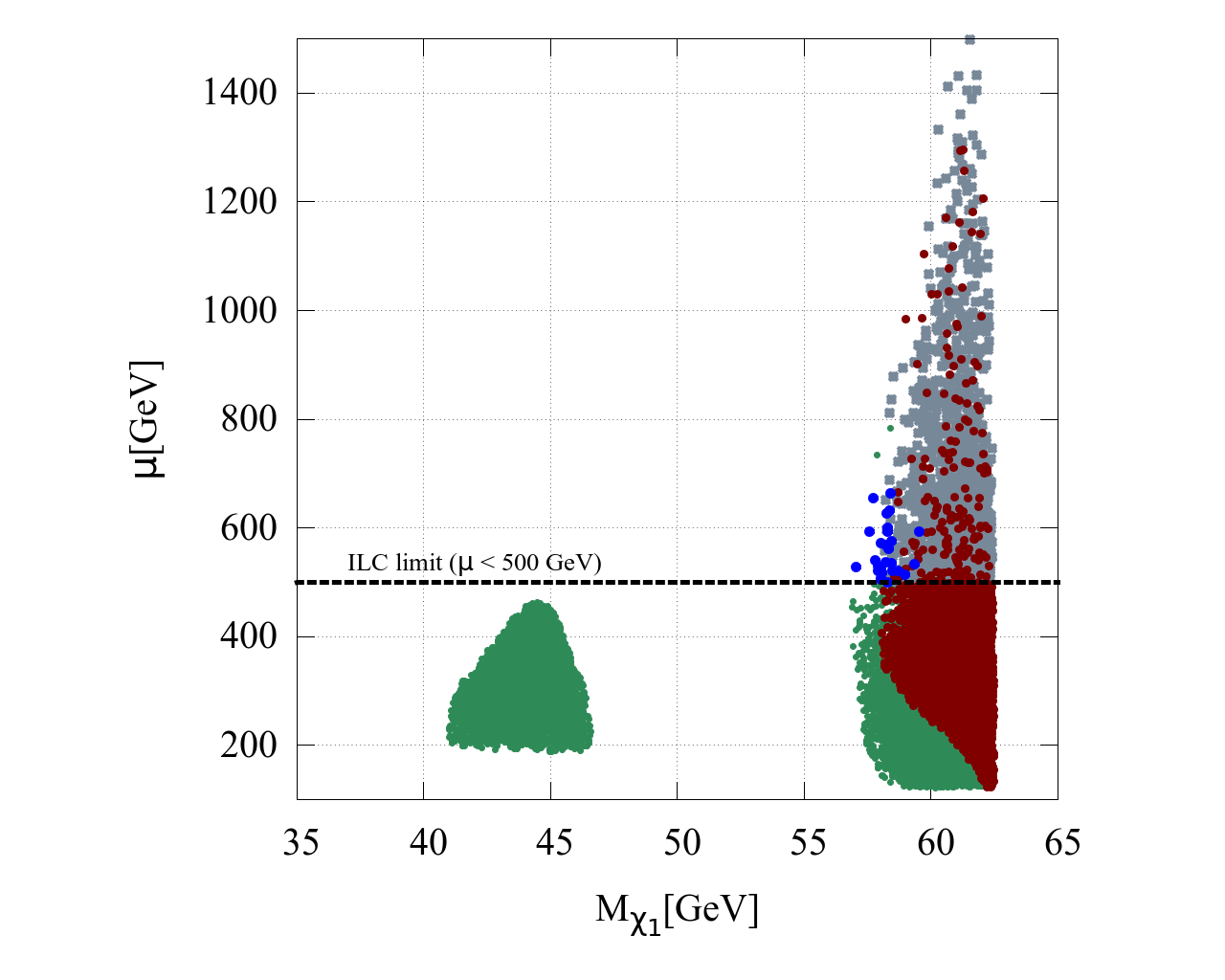}
\caption{ \textbf{a)}  The normalized relic density, $\xi = 
\Omega_{DM}/0.122 $ vs the LSP mass, same color code as in 
Fig.~\ref{Thermal:Br_Inv}(a).
\textbf{b)}  higgsino mass parameter $\mu$ against the LSP mass,  the 
black dashed line represents the ILC sensitivity to probe $\mu~<~500~{\rm 
GeV}$. Here, only the parameter points  allowed by collider constraints 
and  LUX-2016 have been considered. The color code is described in the 
text.}
\label{Thermal:mu_dd}
\end{center}
\end{figure}

We show the variation of $\xi$ with the LSP mass  in 
Fig.~\ref{Thermal:mu_dd}(a). Finally, we comment on the prospects  to 
probe the allowed parameter space at the ILC, in particular through 
chargino searches. In this channel, all kinematically accessible chargino 
pairs can be probed. 
To be conservative  and for simplicity, we define the ILC-1TeV  reach as 
$\mu \;{\rm or} \; M2<500$~GeV, following \cite{Baer:2013vqa}. In  
Fig.~\ref{Thermal:mu_dd}(b), we display all  parameter points  allowed by 
collider, relic density and LUX-2016 constraints  in the $\mu-M_{\lspone}$ 
plane. We distinguish four possible scenarios after considering the 
following two possible modes of DM probe by ILC: 

\begin{itemize}
\item[] \textbf{Mode A} : through Higgs to invisible branching, $Br(h \to \lspone \lspone)~>~0.4\%$
\item[] \textbf{Mode B} : through electroweakino searches : $ \mu,~M_{2} < 500 ~{\rm GeV}$.  
\end{itemize}

\begin{itemize}
\item \textit{Probe via mode A and mode B:}  These points are shown in 
green and have $Br(h \to \lspone \lspone)~>~0.4\%$ as well as $\mu\; {\rm 
or} \; M_2~<~500~{\rm GeV}$. 
\item \textit{Probe via mode A only:} These points, in blue, have  $Br(h 
\to \lspone \lspone)~>~0.4\%$ as well as $\mu, M_2~>~500~{\rm GeV}$.
\item \textit{Probe via mode B only:}  These points, in brown correspond 
to $Br(h \to \lspone \lspone)~<~0.4\%$ and $\mu\; {\rm or} \; 
M_2~<~500~{\rm GeV}$.
\item \textit{Cannot be probed by ILC:}  These points are shown in grey.   
\end{itemize} 

The ILC can therefore completely probe the Z-funnel region, just as would 
be possible with DM direct detection with Xenon-1T, while a fraction of 
the Higgs funnel remains out of reach. We find only a limited number of 
points that can be exclusively probed through the Higgs invisible width.

\begin{figure}[ht]
\begin{center}
 \includegraphics[height=2.42in,width=3.0in]{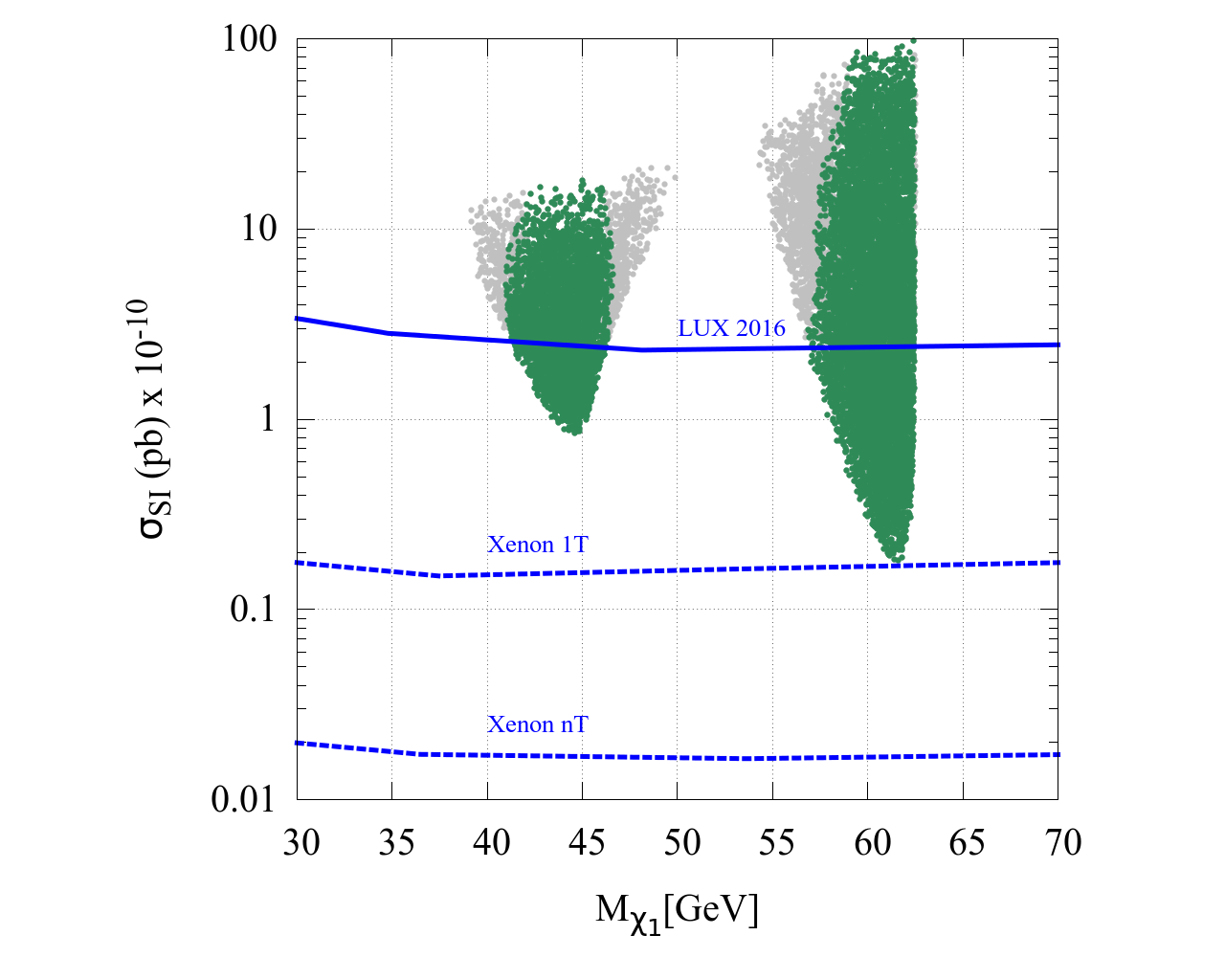}
\caption{Unscaled SI WIMP-nucleon cross-section vs $M_{\lspone}$ for all 
points allowed by collider and relic density constraints. The  blue-solid 
line shows the current limit from LUX-2016~\cite{Szydagis:2016few} and 
the blue-dashed lines show the projected reach for  
Xenon-1T~\cite{Aprile:2015uzo} and Xenon-nT~\cite{Aprile:2015uzo}.}
\label{Thermal:mu_dd_unscale}
\end{center}
\end{figure}

Before concluding this section we comment on the implications of allowing 
for a non-thermal mechanism to increase the value of the relic density to 
the observed value. This entails that the neutralino would constitute all 
of the DM, thus  there is no need to rescale the elastic scattering cross-
sections on nucleons.  
 In Fig.~\ref{Thermal:mu_dd_unscale}, we show the unscaled $\sigma_{SI}$ 
 against $M_{\lspone}$, with the green and grey colored points being the 
 same as that of Fig.~\ref{Thermal:Br_Inv}(b).
 As expected, the constraint from LUX is now more stringent and  a 
 significant number of parameter space points are now excluded. This 
 effect is more prominent in the Z-resonance region, where we observe an 
 upward shift in the funnel region, a factor three improvement over the 
 current LUX limits would suffice to exclude this region. 
 In the next section, we further relax the assumption of a thermal relic 
 and explore the parameter space specified in 
 Sec.~\ref{sec:par_space_scan}, in the context of limits from ILC and DM 
 direct detection experiments, following a similar approach to the one 
 adopted in this section.

\section{Probing overabundant neutralinos}\label{sec:non_standard}

Assuming a thermal production of DM, we have seen that the relic density  
provides strong constraints on the MSSM parameter space.  In particular, 
bino-like neutralinos lighter than roughly 34 GeV are ruled out as their 
annihilation cross-section is too small, leading to a predicted value for 
the relic density ($\Omega_{CDM}$) that can be orders of magnitude larger 
than the observed value ($\Omega_{DM}$). However, there is ample 
motivation for considering non-thermal mechanisms that lead to much 
different predictions and that allow to reproduce the observed value of 
the DM relic density. The prime example is the  case of a late decaying 
particle, such as a SUSY modulus scalar, where, the late decaying particle 
dilutes the entropy density and can thus lead to the correct value for the 
relic density of DM ($\Omega_{\lspone}$). The amount of dilution is 
sensitive to the mass of the heavy decaying particle ($m_\phi$). The 
dilution also depends on the mass of the DM, the branching ratio of the 
heavy particle to DM and the reheating temperature. Hence, for every value 
of DM mass, several combinations of reheating temperature and heavy scalar 
mass can lead to  a relic density compatible with the observed value.

\begin{equation}
\frac{\Omega_{\lspone}}{\Omega_{DM}} \propto b n_\phi m_\chi T_{RH}
\end{equation}
where $b$ is the number of neutralinos produced per $\phi$ decay, $n_\phi$ 
the number density of the scalar and $T_{RH}$  the reheating 
temperature~\cite{Gelmini:2006pw,Gelmini:2006pq}.
Here, we will not perform a detailed investigation of a specific non-
thermal mechanism  but will simply assume that it is possible to find a 
mechanism that brings  the DM relic density in agreement with 
observations. Hence, in practice we will analyse all those parameter space 
points, for which  the relic density values computed assuming thermal 
freeze-out with a standard cosmological model, is above the measured 
value, 
\begin{eqnarray}
\Omega_{\lspone} h^2 ~>~ 0.122\text,
\label{eqn:nt_rd}
\end{eqnarray} 
and we will investigate the characteristics and the signatures of the 
thermally over-abundant neutralino, hereafter called NSDM neutralino.

\begin{figure}[ht]
\begin{center}
\includegraphics[scale=0.17]{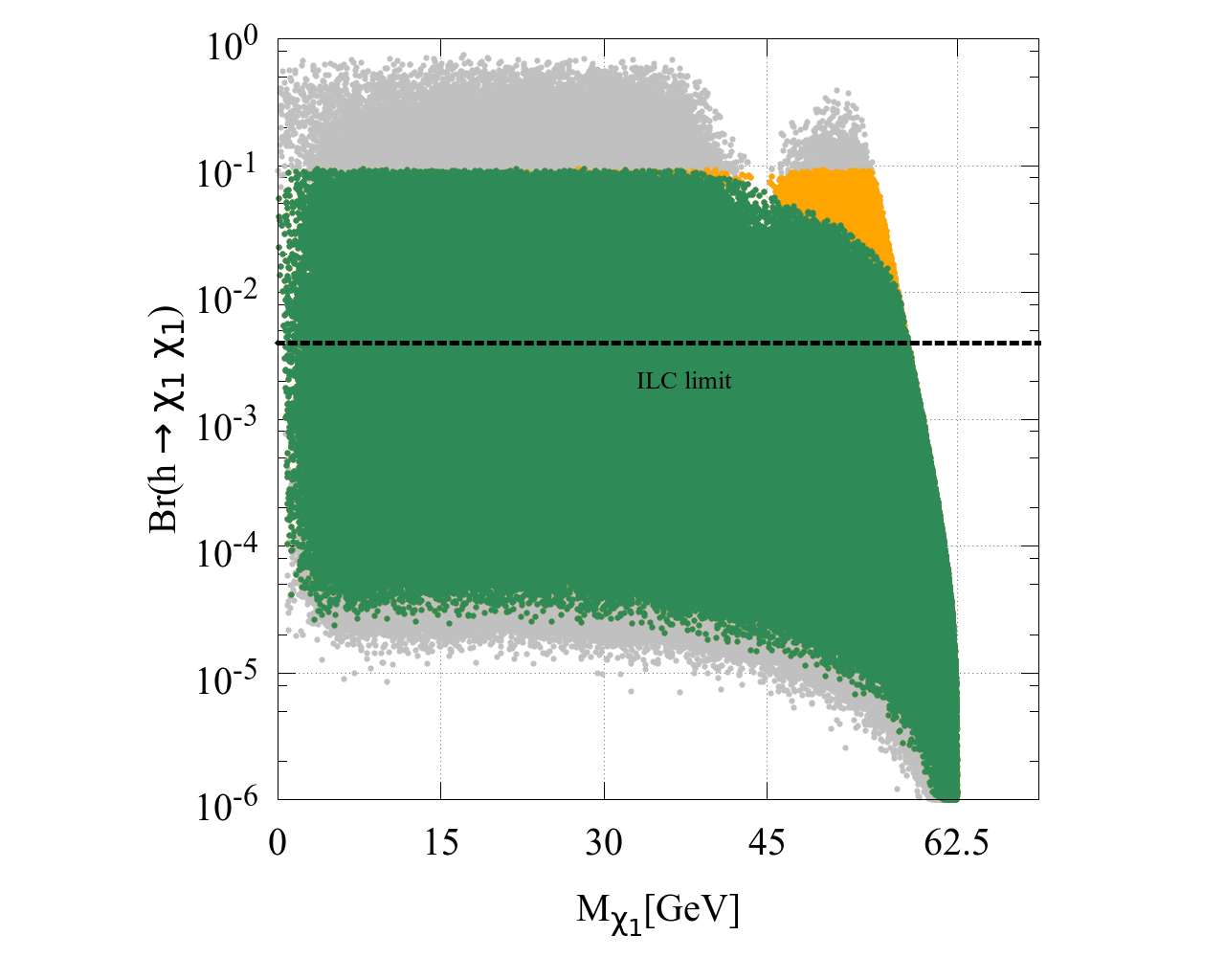} \includegraphics[scale=0.17]{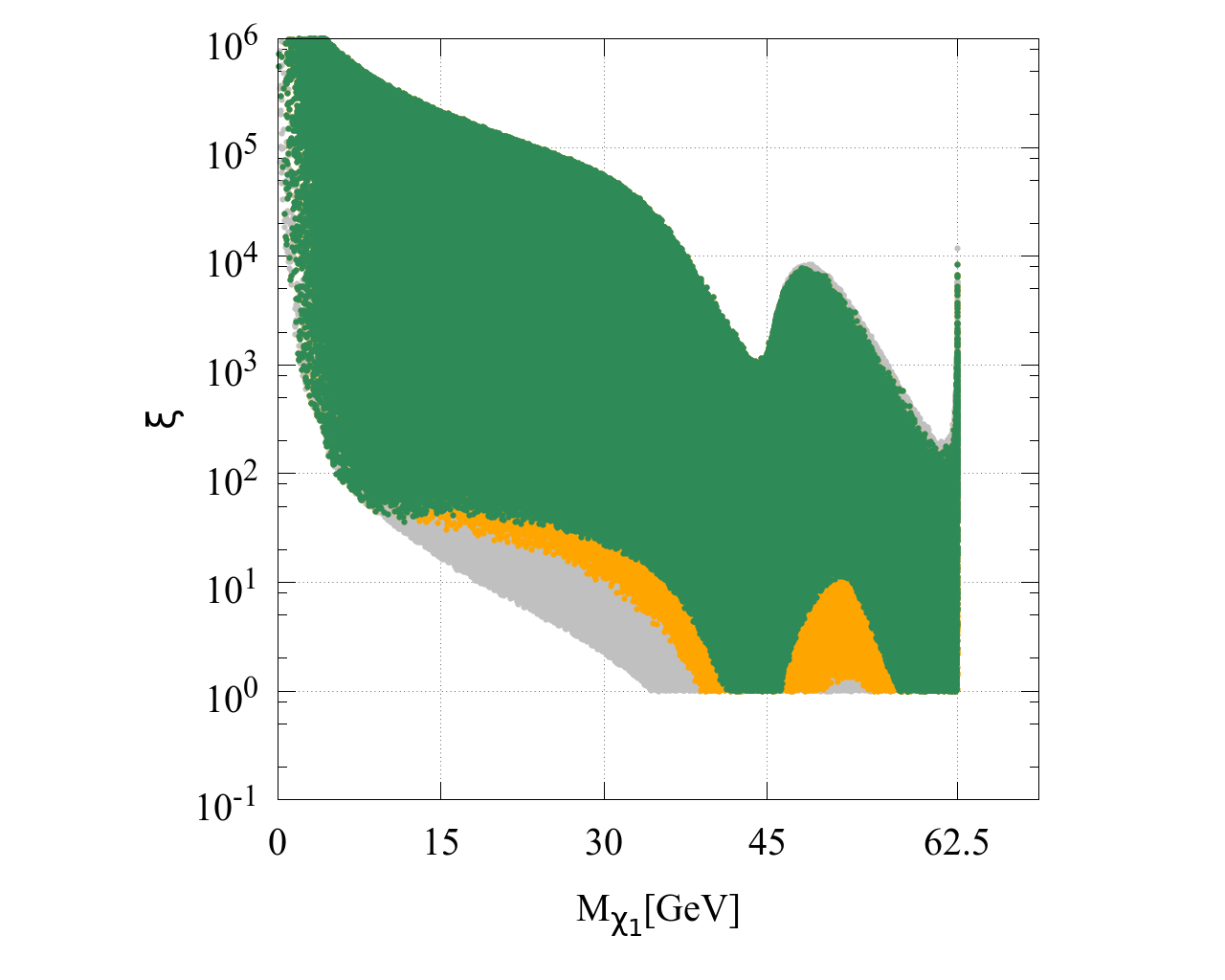}
\caption{\textbf{a)} The Higgs to invisible branching  fraction $Br(h \to 
\lspone \lspone)$ vs. the LSP mass $M_{\lspone}$.  \textbf{b)} The 
rescaled relic density,  $\xi$,    against $M_{\lspone}$. Same colour code 
as Fig.~\ref{Thermal:Br_Inv}.}
\label{NonThermal:Br_Inv}
\end{center}
\end{figure}

With this condition, the lower bound on the lightest neutralino  mass is 
lifted and we obtain LSP's with masses that span the whole range of the 
scan (upto $62.5~{\rm GeV}$). 
The resulting Higgs to invisible branching fraction ($Br(h \to \lspone 
\lspone)$), for all points that satisfy all the constraints mentioned in 
Sec.~\ref{sec:par_space_scan}, have been shown in color in 
Fig.~\ref{NonThermal:Br_Inv}(a). The grey points in the same figure are 
excluded by the Higgs signal strength constraints. As in the previous 
section, we observe that imposition of the Higgs signal strength 
constraints translate into an upper bound on the Higgs to invisible 
branching fraction, which is again approximately $\approx 10\%$. In 
addition, we observe that some parameter space points with a very small  
Higgs to invisible branching fraction,  $Br(h \to \lspone \lspone) 
\lesssim 10^{-6}$, are also  disallowed by the Higgs signal strength 
constraints.
For these points, the partial decay width of $h \to b\bar{b}$ attains a 
value which is appreciably greater than the SM expectations, leading to 
significant deviations from the SM, in the branching of $h \to ZZ$ and $h 
\to WW$ (and consequently $h\to \gamma\gamma$). 
We find that allowing for non-standard cosmology, the Higgs to invisible 
branching fraction  can vary over a wide range and can attain values as 
low as $Br(h \to \lspone \lspone) \sim 10^{-6}$ in the Higgs resonance 
region. 
It can be observed from Fig.~\ref{NonThermal:Br_Inv}(a) that very small 
values of $Br(h \to \lspone \lspone)$ can be obtained for very light DM, 
which is a direct consequence of relaxing the relic density constraint, 
since an efficient annihilation mechanism is no longer required, and thus, 
the coupling of the LSP to the Higgs can be very small. 
The yellow points in Fig.~\ref{NonThermal:Br_Inv}(a) are excluded by the 
current limits on SI WIMP-nucleon interaction cross-sections from 
LUX-2016~\cite{Szydagis:2016few}. Typically, these yellow colored points 
correspond to a large LSP-Higgs coupling, resulting in a large invisible 
width, and hence, excludes the region with $M_{\lspone}\gtrsim 15~{\rm 
GeV}$, where direct detection has a better sensitivity. The green colored 
parameter space points are still allowed by all colliders and direct 
detection limits and will be referred to as the allowed parameter space in 
the remainder of this section. 
Assuming a standard thermal DM scenario, we obtain values of the relic 
density, illustrated by $\xi$ in 
Fig.~\ref{NonThermal:Br_Inv}(b). It can be seen that the value of $\xi$ 
for the allowed parameter space is at least around two orders of magnitude 
above the observed limits for all $M_{\lspone} \lesssim 30~{\rm GeV}$ and 
reach the limit of validity of the micrOMEGAs computation for 
$M_{\lspone}\lesssim 10~{\rm GeV}$ ($\xi\sim10^{6}$).

\begin{figure}[ht]
\begin{center}
\includegraphics[scale=0.25]{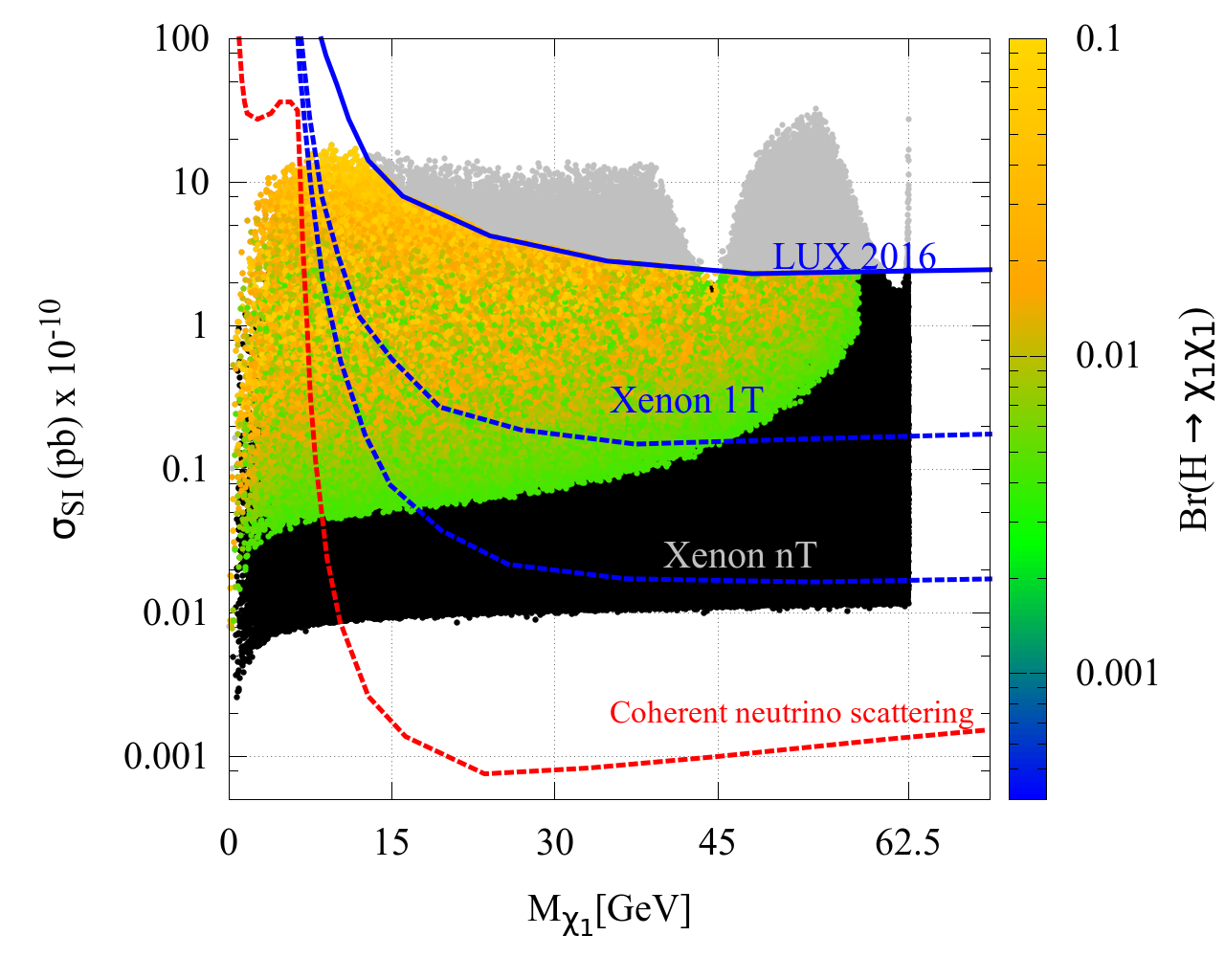}
\caption{SI WIMP-nucleon cross-section vs $M_{\lspone}$ for all points 
allowed by collider and relic density constraints. The color code 
characterizes the value of $Br(h \to \lspone \lspone)$, while black points 
have $Br(h \to \lspone \lspone)<0.4\%$ The  blue-solid line shows the 
current limit from LUX-2016~\cite{Szydagis:2016few} and the blue-dashed 
line shows the reach for  Xenon-1T~\cite{Aprile:2015uzo} and Xenon-
nT~\cite{Aprile:2015uzo}.}
\label{NonThermal:mu_dd_2}
\end{center}
\end{figure}

The prospects for direct detection are illustrated in 
Fig.~\ref{NonThermal:mu_dd_2}, where we show the limits on SI WIMP-nucleon 
cross-section from LUX-2016~\cite{Szydagis:2016few} and the reach of 
Xenon-1T~\cite{Aprile:2015uzo} and Xenon-nT~\cite{Aprile:2015uzo}.
Contrary to the thermal DM case, a significant fraction of the points are 
below the reach of Xenon-1T and even the future Xenon-nT. 
Clearly many points are below the threshold for detection but there are 
also points with $M_{\lspone} > 30~{\rm GeV}$ which will be undetectable 
at the large-scale detectors, where the direct detection (DD) experiments 
have an excellent sensitivity. Moreover the SI cross-section at low mass 
is predicted to lie below the coherent neutrino background. Thus even 
future experiments such as SuperCDMS-SNOLAB~\cite{Agnese:2016cpb} designed 
to enhance the sensitivity at low masses will not be able to probe this 
region.  
 
\begin{figure}[ht]
\begin{center}
\includegraphics[scale=0.18]{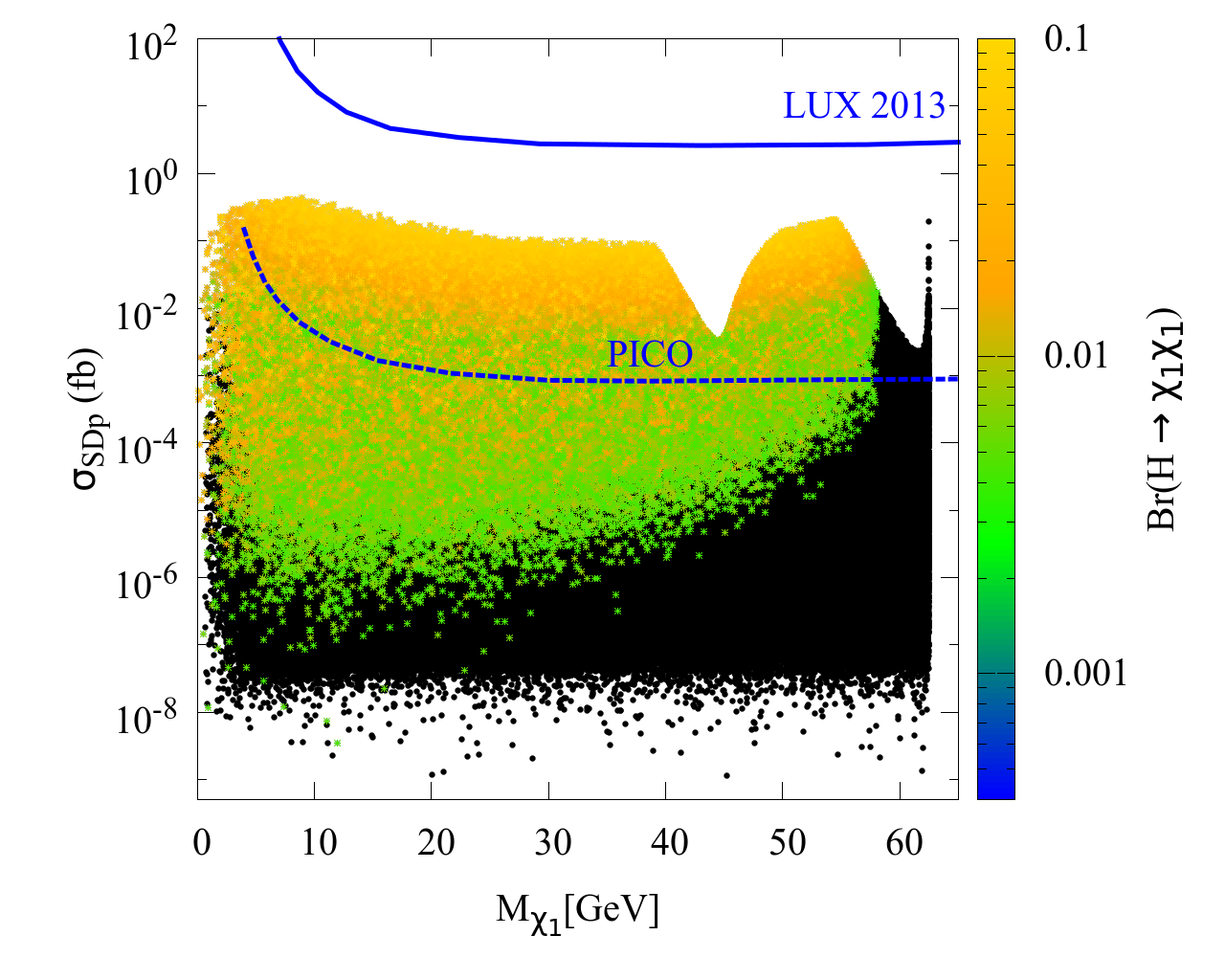}\includegraphics[scale=0.18]{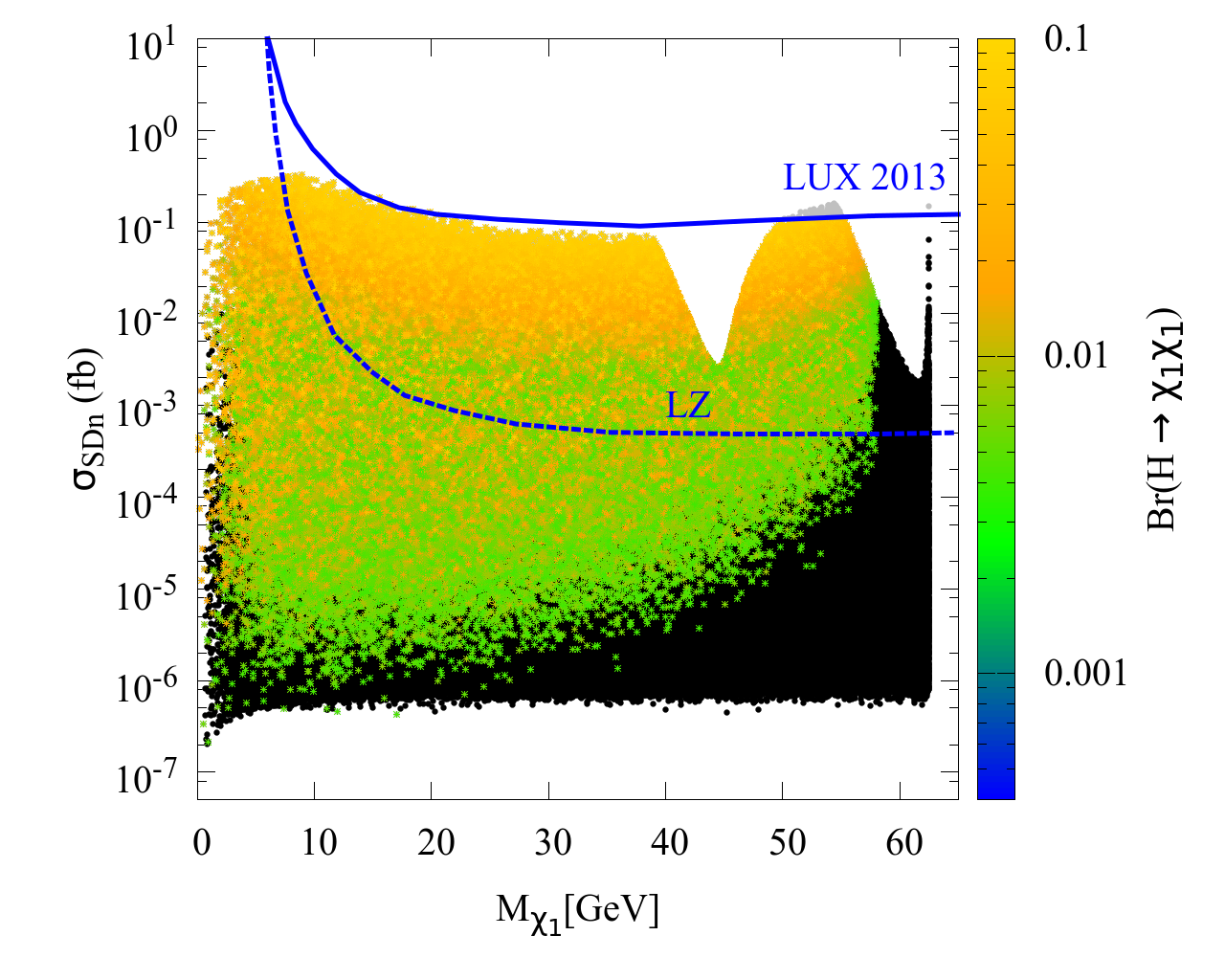}
\caption{\textbf{(a).} SD WIMP-proton cross-section vs $M_{\lspone}$ for 
all points allowed by collider and relic density constraints. The 
blue-solid and blue-dashed line shows the current limits from 
LUX-2013~\cite{Akerib:2016lao} 
and the reach of PICO-250~\cite{Cushman:2013zza}, respectively. 
\textbf{(b).} SD WIMP-
neutron cross-section vs $M_{\lspone}$ for all points allowed by collider 
and relic density constraints. The  blue-solid line and the blue-dashed line shows the current 
limits from LUX-2013~\cite{Akerib:2016lao} and the reach of 
LZ~\cite{Akerib:2016lao}, respectively. The color code characterizes the value of $Br(h 
\to \lspone \lspone)$, black points have $Br(h \to \lspone 
\lspone)<0.4\%$.}
\label{NonThermal:mu_sd_2}
\end{center}
\end{figure}

To draw a better illustration of the complementarity between the Higgs 
invisible branching and the DD cross-sections, we colour-coded the allowed 
points in Fig.~\ref{NonThermal:mu_dd_2} according to the Higgs invisible 
branching fraction. Points with $Br(h \to \lspone \lspone) \lesssim 0.4\%$ 
have been shown in black, while, an overlapping color palette has been 
used for those parameter space points which can be probed at ILC through 
the Higgs to invisible branching fraction.  
 Interestingly, a large fraction of the points that are below the 
 threshold for direct detection are within the reach of the ILC, while  a 
 fraction of points can be probed both at the ILC and with ton scale DD 
 detectors. There also exists such points which are out of reach of both 
 the detection methods. We present the parameter space points which 
 satisfy all the constraints of Sec.~\ref{sec:par_space_scan} in the 
 $\sigma_{SD}^{\lspone prot.}$ - $M_{\lspone}$ ($\sigma_{SD}^{\lspone 
 neut.}$ - $M_{\lspone}$) plane as well, in Fig.~\ref{NonThermal:mu_sd_2}
 (a) (Fig.~\ref{NonThermal:mu_sd_2}(b)). It can be observed that the 
 current limits from LUX-2013 are not strong enough to exclude the 
 parameter space points except for a few at $M_{\lspone}\approx 55~{\rm 
 GeV}$, which are excluded by the SD WIMP-neutron cross-section limits, as 
 evident from Fig.~\ref{NonThermal:mu_sd_2}(b).

\begin{figure}[ht]
\begin{center}
\includegraphics[scale=0.22]{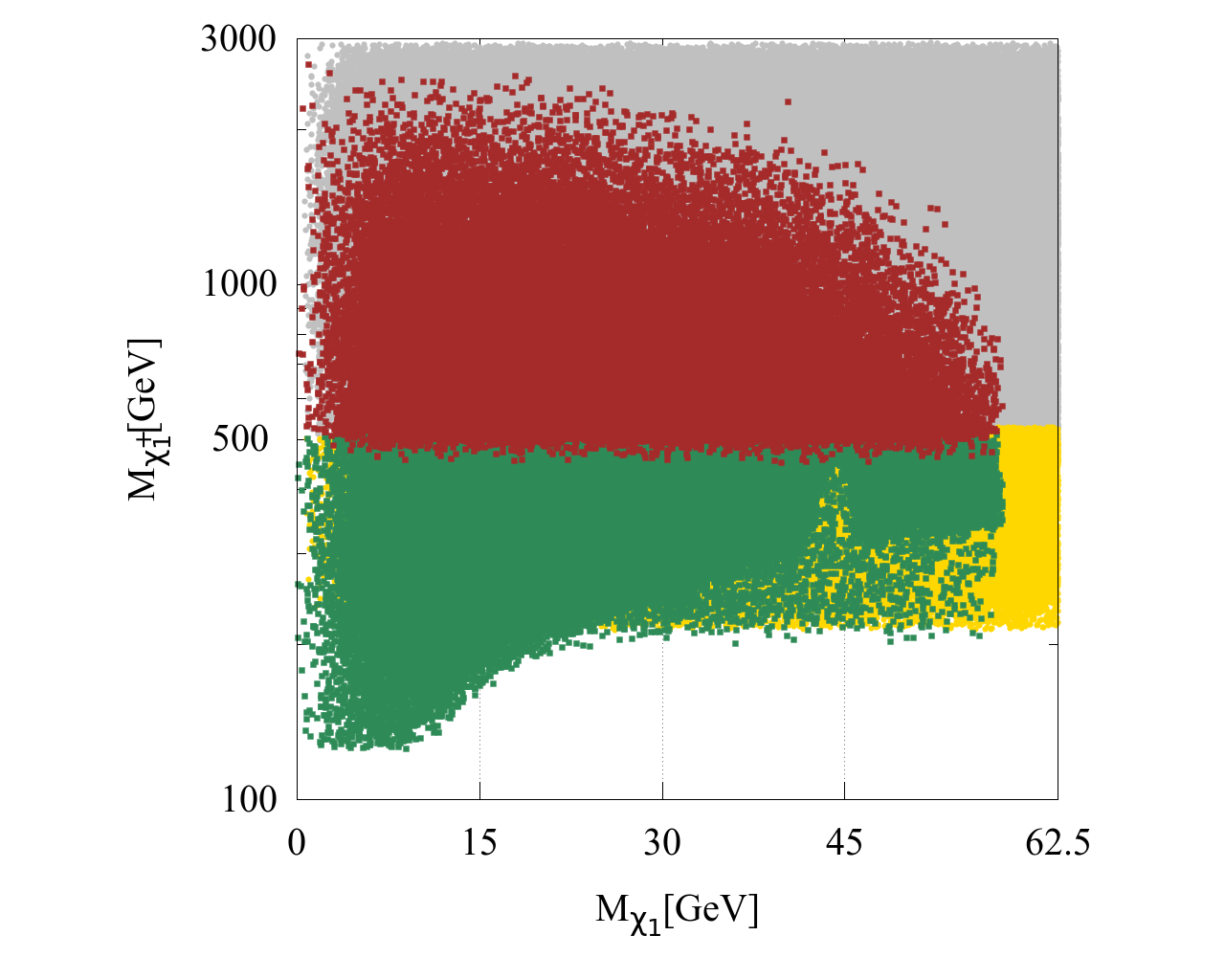} 
\caption{Represents the chargino mass ($M_{\chonepm}$) against $M_{\lspone}$ for the allowed parameter space points. The corresponding color code is mentioned in the text.}
\label{NonThermal:mu_dd}
\end{center}
\end{figure}

We present the allowed parameter space in the $M_{\chonepm} - M_{\lspone}$ 
plane in Fig.~\ref{NonThermal:mu_dd} in order to investigate the 
implications of ILC sensitivity to probe $\mu,M_{2}<500~{\rm GeV}$ through 
electroweakino searches. $\mu~=~500~{\rm GeV}$ has been shown in black-
dashed line in Fig.~\ref{NonThermal:mu_dd}. Similar to the previous 
section, the parameter space points in Fig.~\ref{NonThermal:mu_dd} have 
been classified into following four different categories based on the two 
different detection modes of ILC,
\begin{itemize}
\item[] \textbf{Mode A} : through Higgs to invisible branching, $Br(h \to 
\lspone \lspone) > 0.4\%$  
\item[] \textbf{Mode B} : through electroweakino searches, $ \mu,M_{2} < 
500 ~{\rm GeV}$.  
\end{itemize}

\begin{itemize}
\item \textit{Probed by ILC through mode A and mode B:} These points have 
been represented in green color in Fig.~\ref{NonThermal:mu_dd}. These 
parameter space points have $Br(h \to \lspone \lspone)~>~0.4\%$ as well as 
either $\mu~<~500~{\rm GeV}$ and/or $M_{2}~<~500~{\rm GeV}$.
\item \textit{Probed by ILC only through mode A:} These parameter points 
have been represented in brown color. Both, $\mu$ and $M_{2}$ is above 500 
GeV for these points, and $Br(h \to \lspone \lspone) > 0.4\%$.
\item \textit{Probed by ILC through mode B:} We show these parameter space 
points in yellow color. These parameter points have $\mu~<~500~{\rm GeV}$ 
and/or $M_{2}~<~500~{\rm GeV}$, along with, $Br(h \to \lspone \lspone) 
\leq 0.4\%$.
\item \textit{Points which cannot be probed by ILC}: These parameter space 
points have been shown in grey. For these parameter space points, $Br(h 
\to \lspone \lspone) < 0.4\%$, $M_{2}~<~500~{\rm GeV}$ and $\mu~<~500~{\rm 
GeV}$. Hence, these parameter points evade detection by ILC.
\end{itemize}

It can be observed from Fig.~\ref{NonThermal:mu_dd} that ILC will be 
capable of probing a significant fraction of the allowed parameter space 
considering ILC's detection capability through both, mode A and mode B. 
However, a notable fraction of parameter space points will evade detection 
by ILC as well as from the current DM direct detection experiments, as 
evident from the grey colored points in Fig.~\ref{NonThermal:mu_dd}.

\section{Complementarity of future experiments in probing dark matter}\label{sec:comp}

In this section we characterize the different possibilities to identify 
the nature of dark matter by exploiting the complementarity between  the 
ILC, through a precise measurement of the Higgs invisible width or the 
detection of electroweakinos, and  future SI or SD direct detection 
experiments. Here we mean the detectors beyond the ones currently in 
operation, more specifically XENON-nT (through SI WIMP-nucleon based 
interaction), PICO-250 (through SD WIMP-proton based interaction) and LZ 
(through SD WIMP-neutron based interaction). For the sake of simplifying 
the discussion,  we adopt the simple criteria that the limit of 
detectability for electroweakinos at the ILC  is $\mu<500$~GeV.
\footnote{The ILC can also just as easily probe values of $M_2<500$~GeV, 
however the nearly pure winos would be first discovered at the LHC. }
In addition to detecting new particles,  the ILC will also be capable of 
performing very precise electroweakino mass measurements with an 
uncertainty of less than 1 GeV~\cite{Li:2010mq} and thus will allow to  
determine the gaugino masses and the value of $\mu$.  Moreover we note 
that  the LSP mass  can be  determined in direct detection experiments 
albeit with a large uncertainty. This also requires  that  a certain 
number of events are observed~ \cite{Drees:2008bv}. For example the mass 
of a WIMP of  $\sim 50 ~{\rm GeV}$ can be measured to $\sim 35\%$ with 100 
events.
 To organize the discussion, in the following subsections  we group  the 
 points according to the type of experiment that have the potential to 
 probe them. For this we consider all scenarios that satisfy current 
 collider and flavor  constraints regardless of whether thermal dark 
 matter can reproduce the observed relic density and point out the 
 conditions for distinguishing thermal and NSDM scenarios.

\subsection{Detection at the ILC only}

\begin{figure}[htb!]
\begin{center}
\includegraphics[scale=0.16]{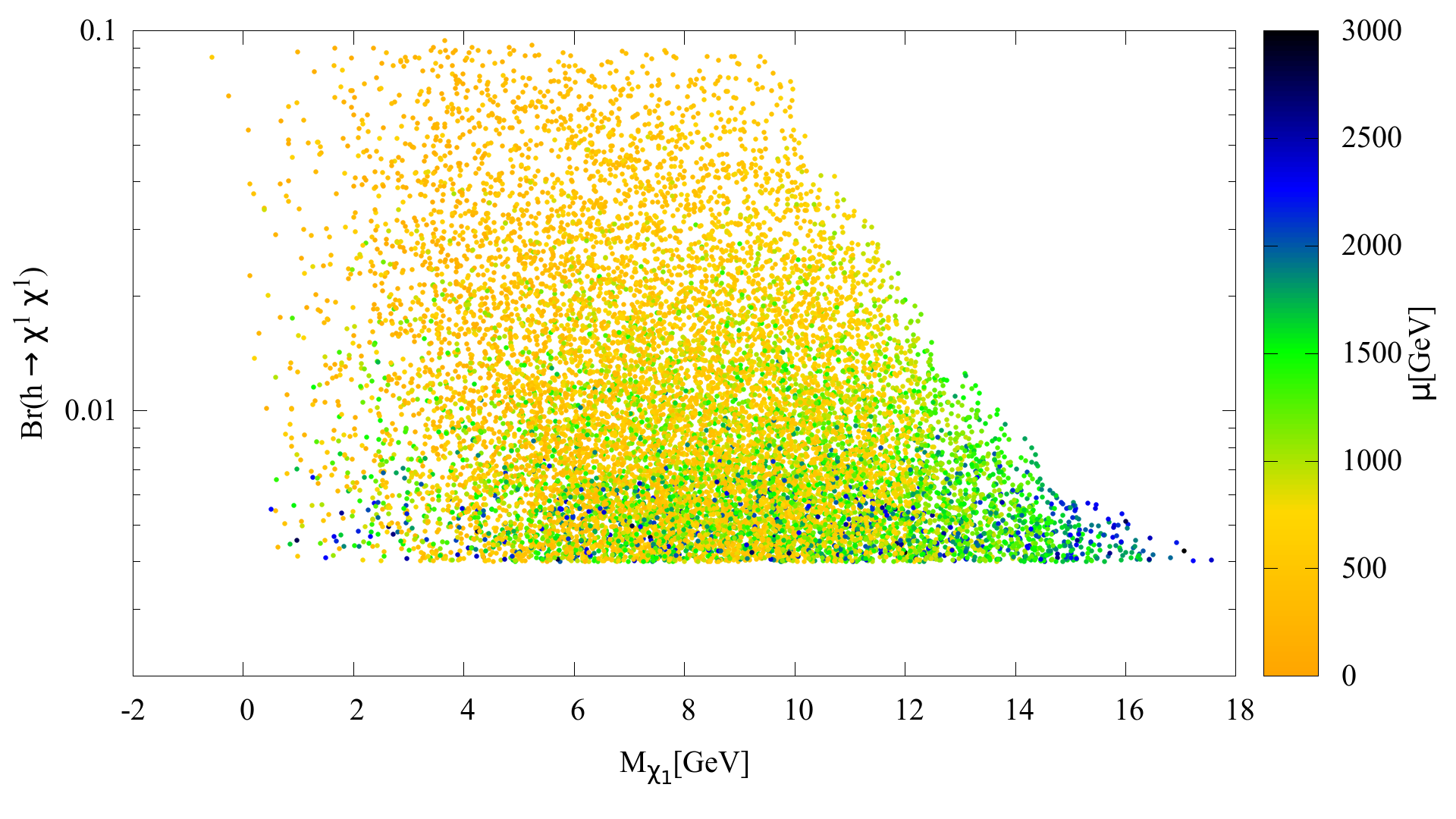}
\caption{Scatter plot in the $Br( h \to \lspone \lspone) - M_{\lspone}$ 
plane for parameter space points which can be probed by ILC only, through 
the Higgs to invisible branching fraction. The color palette corresponds 
to the value of the higgsino mass parameter ($\mu$)}
\label{comp_ILC_Br}
\end{center}
\end{figure}

The first class of scenarios we consider are those that can be probed 
exclusively  through the Higgs to invisible branching fraction at ILC. 
These parameter points will evade detection at all the future DD 
experiments (Xenon-nT, PICO-250 and LZ) considered in this analysis. We 
show the parameter points with  $Br(h \to \lspone \lspone)~>~0.4\%$ in the 
$Br( h \to \lspone \lspone) - M_{\lspone}$ plane in 
Fig.~\ref{comp_ILC_Br}. The color palette  corresponds to the value of 
$\mu$.

\begin{figure}[htb!]
\begin{center}
\includegraphics[scale=0.16]{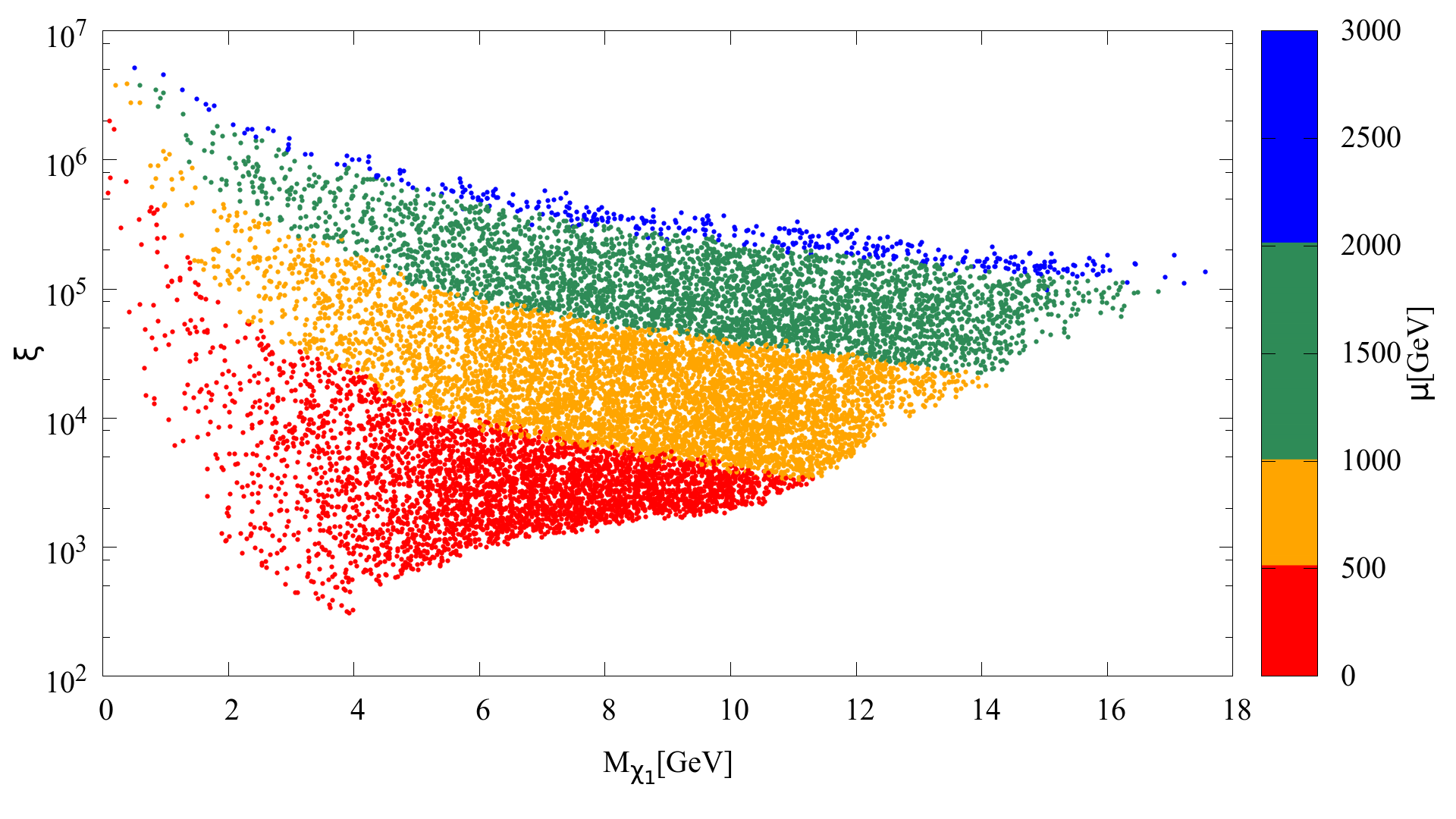}
\caption{Scatter plot in the $\xi - M_{\lspone}$ plane for parameter space 
points which can be probed by ILC only, through the Higgs to invisible 
branching fraction ($Br(H \to \lspone \lspone) \geq 0.4\%$). The color 
palette corresponds to the value of the higgsino mass parameter ($\mu$)}
\label{comp_ILC_xi}
\end{center}
\end{figure}

As expected the points are concentrated in the low mass region, 
$M_{\lspone}\lesssim 18~{\rm GeV}$, corresponding to DM masses mostly 
inaccessible to direct detection. 
 In this case, the branching fraction of $h \to \lspone \lspone$ goes  up 
 to  $\approx 10\%$  for relatively low values of $\mu$ as  shown in the 
 color palette in Fig.~\ref{comp_ILC_Br}. 
 Note that the points with  $M_{\lspone}> 12~{\rm GeV}$ are associated 
 with a larger value of $\mu$, thus are more weakly coupled to the Higgs 
 and  evade direct detection limits despite being above the threshold.

 In Fig.~\ref{comp_ILC_xi}, we show the same parameter space points in the 
 $\xi - M_{\lspone}$ plane with the color palette representing the value 
 of $\mu$.
All the points have $\xi> 500$  with the higher values attained at low 
$\lspone$ masses.
It can be concluded from Fig.~\ref{comp_ILC_xi} that observation of a DM 
signal exclusively  through the Higgs to invisible branching fraction, 
would be a strong indication for the DM candidate to be light and an 
artifact of non-standard cosmology. 

A subclass of the scenarios that can be probed through Higgs invisible at 
the ILC feature $\mu \leq 500~{\rm GeV}$, and are thus 
accessible through electroweakino searches at ILC. We observe that  such 
scenarios are restricted to $M_{\lspone}\lesssim 12~{\rm GeV}$ and that a 
combined determination of the LSP mass and of $\mu$ at the ILC would  
clearly point towards non thermal scenarios since $\xi \sim [2500:5000]$. 

A similar conclusion can be reached for another subclass of scenarios 
which evade future DD limits and are detectable exclusively at ILC through 
the electroweakino searches. These scenarios feature  a smaller value of  
$M_1$, leading to $M_{\lspone}\lesssim 10~{\rm GeV}$, are nearly pure bino 
and thus are associated with $Br(H \to \lspone \lspone) \leq 0.4\%$ even 
though the higgsino parameter is small, $\mu \leq 500~{\rm GeV}$.
 The precise mass determination capability of the  ILC will thus allow to 
 clearly identify NSDM scenarios, since  in this case $\xi$ is found to be  
 $10^{3} - 10^{4}$ for $M_{\lspone} > 5~{\rm GeV}$.  
Note that for the scenarios described in this subsection we have also 
checked the impact of a precise measurement of the Higgs total width at 
the ILC and found that it did not provide  any additional constraints on 
the parameter space.

\subsection{Detection at Xenon-nT only}
\label{sec:xenon}

\begin{figure}[htb!]
\begin{center}
\includegraphics[scale=0.16]{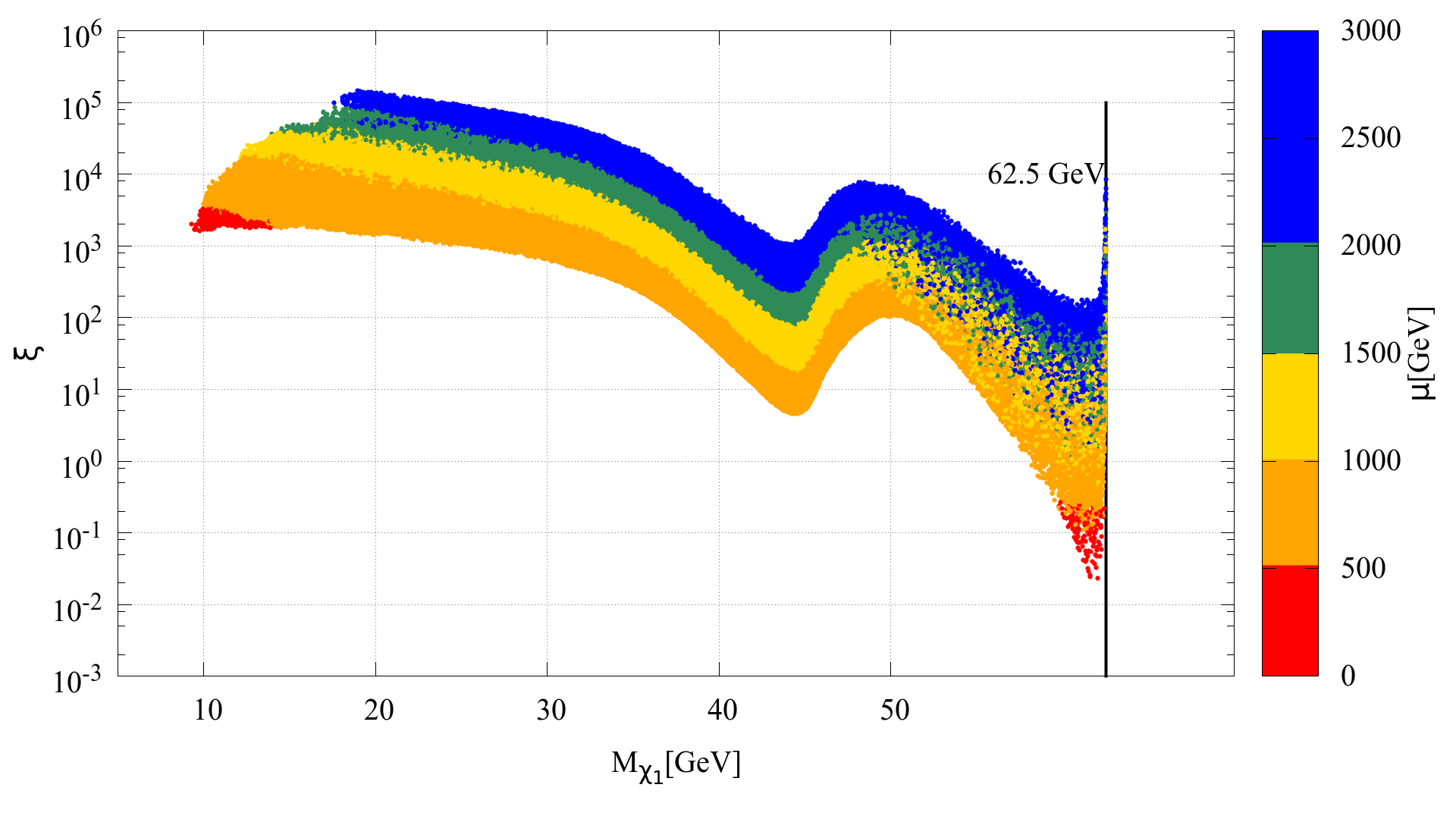}
\caption{Scatter plot in the $\xi - M_{\lspone}$ plane for parameter space 
points which can be probed by Xenon-nT only, through the SI WIMP-nucleon 
interactions. The color palette corresponds to the value of the higgsino 
mass parameter ($\mu$). }
\label{comp_XenT_xi}
\end{center}
\end{figure}

In this subsection, we examine those parameter space points which can be 
probed by the Xenon-nT detector only, through the SI WIMP-nucleon 
interaction cross-sections. 
 These parameter space points evade detection by other future DD 
 experiments, such as PICO-250 and LZ, as well as from 
 ILC through Higgs to invisible branching fraction since $Br( h \to 
 \lspone \lspone)\leq 0.4\% $. We show these parameter space points in 
 Fig.~\ref{comp_XenT_xi} in the $\xi - M_{\lspone}$ plane with $\mu$ 
 represented as a color palette. 

Within this scenario, $\xi$ varies from $\sim 10^{-3}$  all the way up to 
$ \sim 5 \times 10^{4}$ with a spread over 7 orders of magnitude. We 
obtain parameter space points with a  relic density $\Omega_{DM} h^2 < 
0.122$, representing a relic from standard cosmological history, only in 
the Higgs resonance region. 
It is however difficult not only to know precisely enough  the LSP mass to 
establish that it lies within the Higgs resonance region but also to 
identify whether the signature corresponds to a relic from standard or 
non-standard cosmology.

However, there are unique scenarios, where it becomes possible to obtain 
more precise information about the relic density of $\lspone$, based on 
specific categorization of $\mu$ and $M_{\lspone}$. For example, we 
encounter some parameter space points with $\mu<500~{\rm GeV}$, making 
them accessible to ILC through the electroweakino searches. These 
parameter points (shown in red color in   Fig.~\ref{comp_XenT_xi}) are 
restricted to two well separated and compact regions: a) $M_{\lspone} \sim 
9 - 15 ~{\rm GeV}$ with $\xi$ within the range $2500-3000$ and
b. $M_{\lspone} \sim 57 - 62.5 ~{\rm GeV}$ with $\xi$ within the range 
$0.01-0.5$. Taking into account ILC's capability to precisely measure the 
mass of the neutralino,  it can be concluded that:
\begin{itemize} 
\item observation of a signal at Xenon-nT (through SI WIMP-nucleon 
interaction) and at the  ILC (through electroweakino searches only) with a 
LSP compatible with the Higgs resonance region, would  indicate that the 
DM candidate ($\lspone$) could be a relic from standard thermal history.

\item a similar observation, however at low $M_{\lspone} \sim 9 -15~{\rm 
GeV} $, would clearly indicate that the DM candidate is a relic of non-
standard cosmology. 
\end{itemize}

\begin{figure}[htb!]
\begin{center}
\includegraphics[scale=0.16]{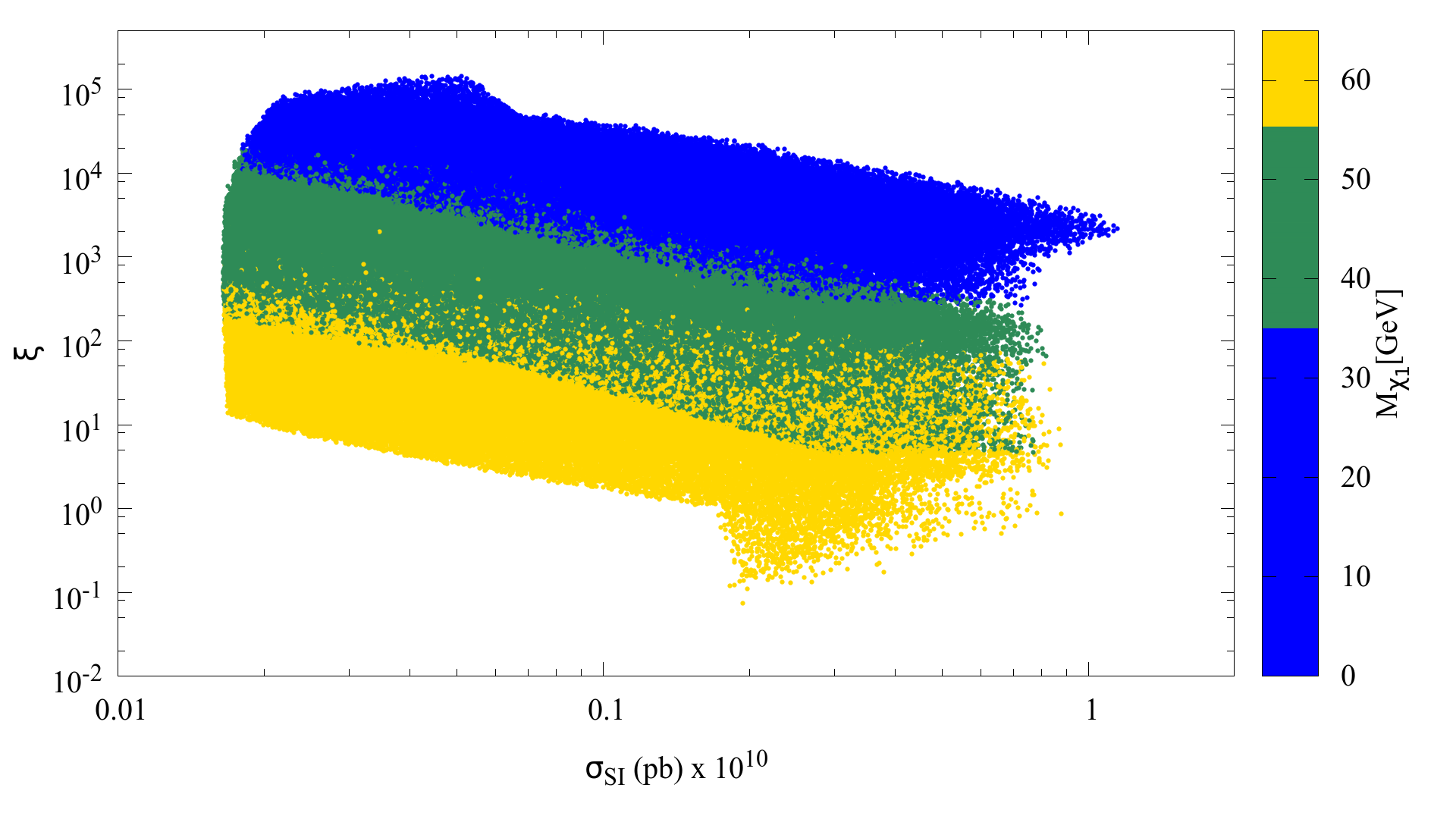}
\caption{Scatter plot in the $\xi - M_{\lspone}$ plane for parameter space 
points which can be probed by Xenon-nT only, through the SD WIMP-nucleon 
interactions. For these parameter space points $\mu > 500~{\rm GeV}$. The 
color palette corresponds to the value of $M_{\lspone}$. }
\label{comp_XenT_xi_mass}
\end{center}
\end{figure}

DM mass measurements carried out by the DD based experiments, even if 
plagued by a large uncertainty  will still be an useful tool in making 
better predictions for $\xi$. In Fig.~\ref{comp_XenT_xi_mass},  we show 
those parameter points of Fig.~\ref{comp_XenT_xi} which have $\mu > 
500~{\rm GeV}$, in the $\sigma_{SI} - \xi$ plane with $M_{\lspone}$ 
represented through the color palette. The ILC will be blind to these 
points in the electroweakino searches. In Fig.~\ref{comp_XenT_xi_mass}, we 
divide the parameter space into three different regions based on 
$M_{\lspone}$. The non-resonant region ($M_{\lspone} = 0-35~{\rm GeV}$) in 
blue, the Z-resonance region ($M_{\lspone} = 35-55~{\rm GeV}$)  in green 
and the Higgs resonant region ($M_{\lspone} > 55~{\rm GeV}$) in yellow. 
In summary, the  Xenon-nT detector offers the best sensitivity for 
$M_{DM}\gtrsim20~{\rm GeV}$. For these parameter space points, $\xi$ 
ranges from $\sim 10^{-2} - \sim 5 \times 10^{5}$ with the highest value 
attained at low $M_{DM}$ and the lowest value attained in the Higgs 
resonance region. In addition, there exists a small subgroup of parameter 
space points with $\mu<500~{\rm GeV}$ which will be sensitive to the 
electroweakino searches at ILC. We found two such regions at very 
different mass ranges,  one for $M_{DM} \sim 9-15~{\rm GeV}$ with NSDM and 
the other 
 for $M_{DM}\sim 57-62.5~{\rm GeV}$ where thermal DM is under-abundant.  
 With the ILC's precise determination of the mass of $\lspone$, an 
 observation under this particular scenario will directly reveal whether 
 the DM candidate is a thermal relic or an outcome of non-standard 
 cosmology. For parameter points with $\mu>500~{\rm GeV}$, the DD 
 experiments can be employed for the determination of the DM masses. 
 Although these mass measurements will have a significant error, they can 
 be useful in inferring the thermal or non-thermal nature of DM, as shown 
 in Fig.~\ref{comp_XenT_xi_mass}.

\subsection{Detection at Xenon-nT and at the ILC with invisible Higgs}

\begin{figure}[htb!]
\begin{center}
\includegraphics[scale=0.2]{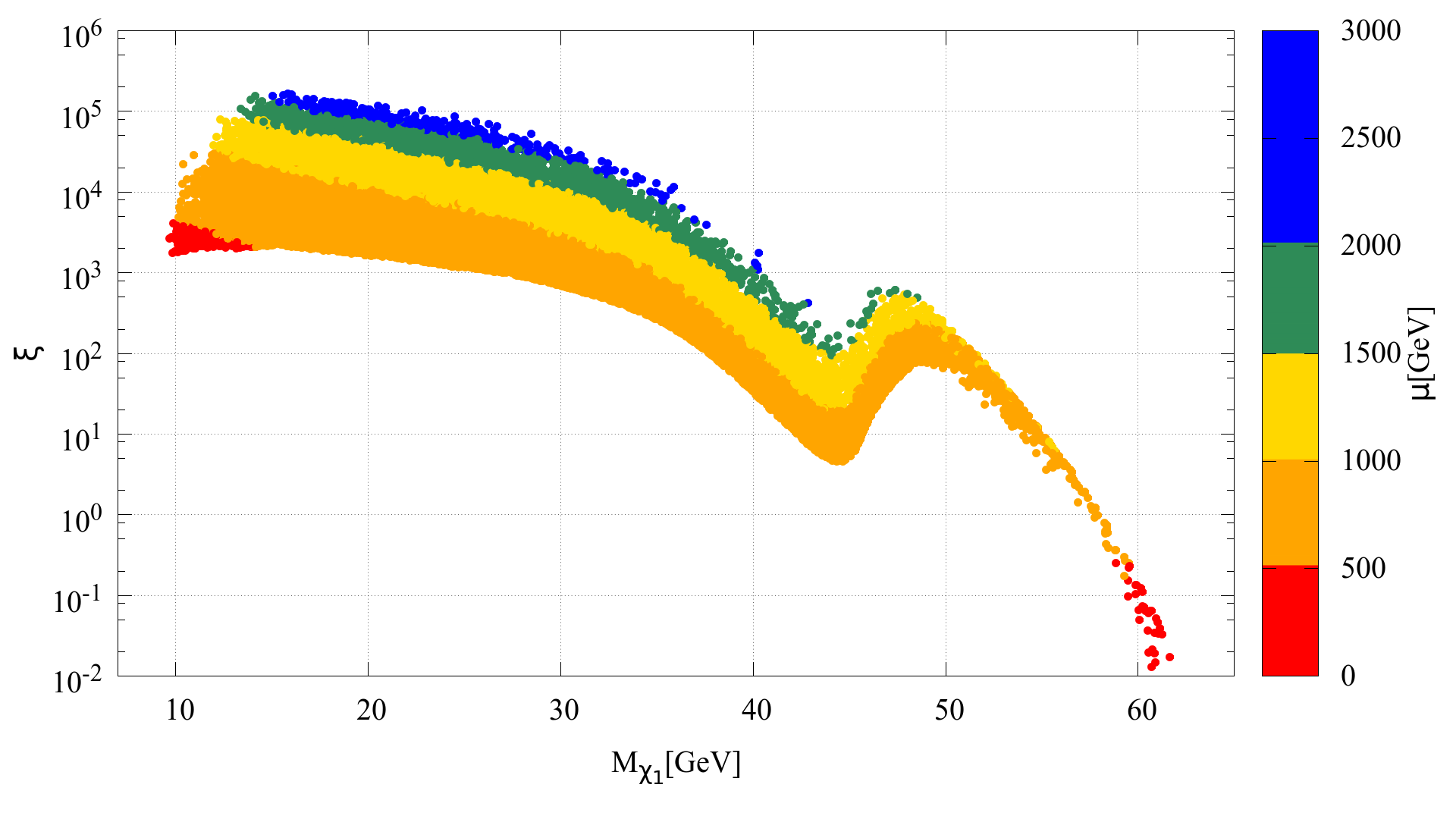}
\caption{Scatter plot in the $\xi - M_{\lspone}$ plane for parameter space 
points points which can be probed by Xenon-nT through the SI WIMP-nucleon 
interactions and also by ILC through the Higgs to invisible branching. The 
color palette corresponds to the value of the higgsino mass parameter 
($\mu$).}
\label{comp_XenT_ilc_xi}
\end{center}
\end{figure}

Here  we  analyse those parameter space points which would be visible at 
ILC through the Higgs to invisible branching fraction and in some cases 
through  electroweakino searches as well and also accessible at Xenon-nT 
through SI based interactions. Other DD experiments considered in this 
analysis would be blind to these parameter space points. We display these 
parameter space points in Fig.~\ref{comp_XenT_ilc_xi}, in the  $\xi - 
M_{\lspone}$ plane along with $\mu$, shown through a color palette.

Parameter space points which fall under the purview of this scenario have 
similar characteristics to those described in the previous section. They 
extend over the mass range $M_{\lspone} \sim 10 - 62~{\rm GeV}$ and are 
associated with either thermal or NSDM scenarios. The only difference with 
the scenario in Sec.~\ref{sec:xenon} is that the LSP must couple 
sufficiently to the Higgs and therefore if close enough to $m_h/2$ will 
lead to $\xi\le 1$.
As above, we observe some parameter space points with $\mu \leq 500~{\rm 
GeV}$ that could be detected through the electroweakino searches at ILC 
(shown in red color in Fig.~\ref{comp_XenT_ilc_xi}). Those are
confined within two separate regions of $M_{\lspone}$, the first with 
$M_{\lspone} \lesssim 15~{\rm GeV}$ and $\xi>1000$, the second  with 
$M_{\lspone}\gtrsim 58~{\rm GeV}$ and  $\xi$ within the range $\sim 0.05 - 
0.5$. With the possibility of a precise measurement of $M_{\lspone}$,   
the following conclusion can be drawn:
\begin{itemize}
\item Observation of a signature at Xenon-nT through the SI WIMP-nucleon 
based interactions and also at ILC through both the Higgs to invisible 
branching fraction and also through electroweakino searches, would 
indicate that the $\lspone$ could be a standard thermal relic, provided 
$M_{\lspone}\gtrsim 58~{\rm GeV}$. 
\item A similar observation, however at low $M_{\lspone}\lesssim 15~{\rm 
GeV}$, would be an indicator of the DM candidate being a relic from non-
standard cosmology.
\end{itemize}

\subsection{Detection with PICO-250, the ILC and/or Xenon-nT}

\begin{figure}[htb!]
\begin{center}
\includegraphics[scale=0.25]{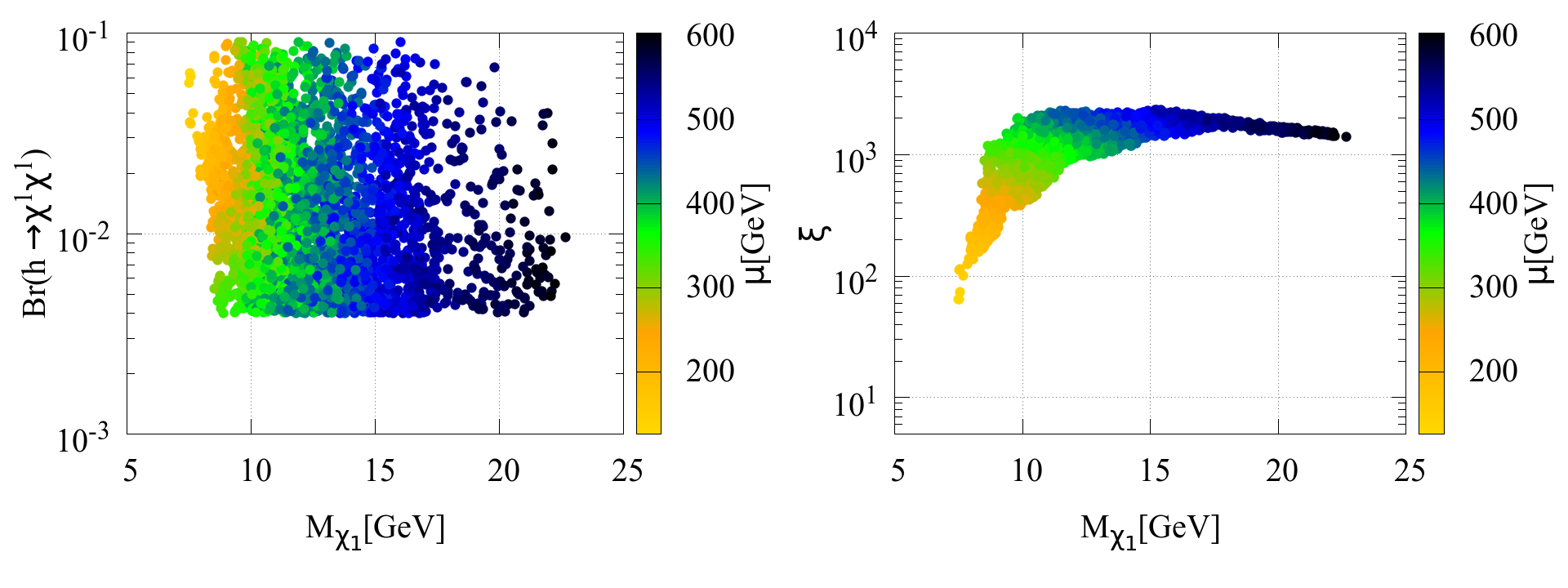}
\caption{(a) Scatter plot in the $Br(h \to \lspone \lspone) - M_{\lspone}$ 
plane and (b) in the $\xi - M_{\lspone}$ plane for parameter space points 
which can be probed by PICO-250 through the SD WIMP-proton interactions, 
by Xenon-nT through the SI WIMP-nucleon based interactions and also by ILC 
through the Higgs to invisible branching. The color palette corresponds to 
the value of the higgsino mass parameter ($\mu$).}
\label{comp_pico_ilc_xent_cat}
\end{center}
\end{figure}

In this subsection, we first analyse those parameter space points which 
can be probed by ILC through the Higgs to invisible branching, by PICO-250 
through the SD WIMP-proton based interactions and also by Xenon-nT through 
the SI WIMP-nucleon based interactions. These parameter space points are 
shown  in Fig.~\ref{comp_pico_ilc_xent_cat}(a) and 
Fig.~\ref{comp_pico_ilc_xent_cat}(b), in the $Br(h \to \lspone \lspone) - 
M_{\lspone}$ and $\xi - M_{\lspone}$ plane, respectively.

We observe that the invisible branching fraction can reach $\sim 10\%$ for 
a LSP in the range  $M_{\lspone}\sim 7 - 23~{\rm GeV}$.  These scenarios 
are associated with NSDM with $\xi$  in the range $ \sim 100 - 3000$, as 
can be seen from Fig.~\ref{comp_pico_ilc_xent_cat}(b) with the low values 
of $\xi$ being attained at low values of $M_{\lspone}$.
Typically  $\mu\leq 500~{\rm GeV}$ to ensure a large enough coupling of 
the LSP to the Z and thus a detectable SD rate which is dominated by Z 
exchange. The value of $\mu$ can however exceed $500$~GeV when  the LSP 
mass increases, $M_{\lspone} \gtrsim 18~{\rm GeV}$, since the higgsino 
fraction of the LSP which drives its coupling to the Z and Higgs is 
determined by $M_1$ and $\mu$.
Moreover  the  sensitivity of detectors also increase with the DM mass. 
For the parameter space points with $\mu\leq 500~{\rm GeV}$, one could 
take advantage of the LSP mass determination at the ILC 
to  clearly indicate that the DM candidate is a relic of non-standard 
cosmology. 

\begin{figure}[htb!]
\begin{center}
\includegraphics[scale=0.10]{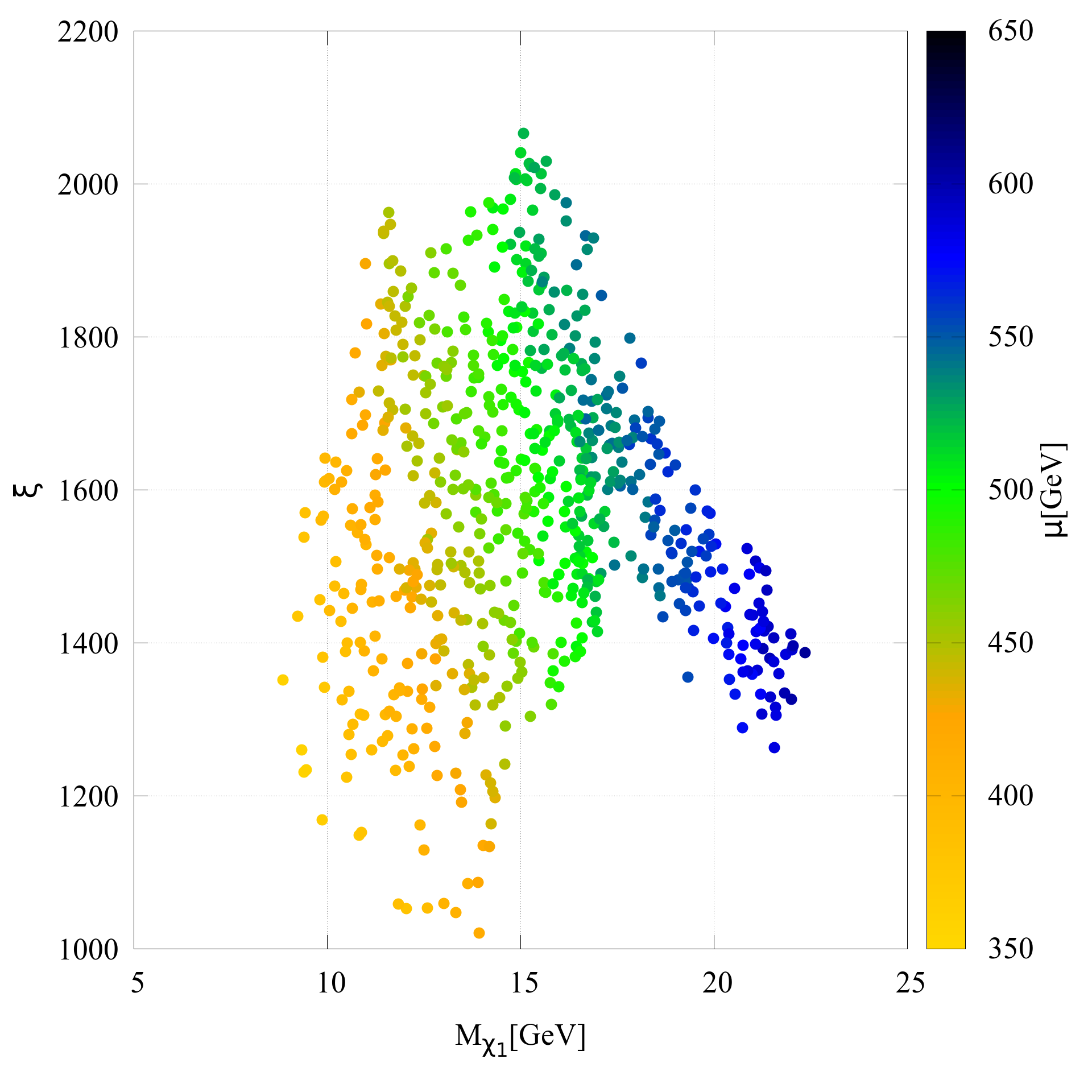}
\caption{Scatter plot in the $\xi - M_{\lspone}$ plane for parameter space 
points which can be probed by PICO-250 through the SD WIMP-proton 
interactions and by Xenon-nT through the SI WIMP-nucleon based 
interactions The color palette corresponds to the value of the higgsino 
mass parameter ($\mu$).}
\label{comp_pico_xent_cat}
\end{center}
\end{figure}

In a similar mass region for the LSP, we observe another interesting 
subgroup of parameter space points,  which would be  accessible to the 
future direct detection experiments only, namely, PICO-250 through the SD 
WIMP-proton based interactions and Xenon-nT through the SI WIMP-nucleon 
based interactions. We show these parameter space points in 
Fig.~\ref{comp_pico_xent_cat}, in the $\xi - M_{\lspone}$ plane. 
 For these points, the Higgs invisible width is too small to be measured 
 because of a slightly higher value for $\mu$. Still only a 
 small fraction of these  points  correspond to $\mu>500~{\rm GeV}$, while 
 most   can be probed by ILC through the electroweakino searches.  A 
 detection in the Xenon-nT detector through the SI WIMP-nucleon based 
 interaction and in PICO-250 through the SD WIMP-proton based interaction, 
 complemented by detection in ILC  through the electroweakino searches, 
 for low $M_{DM}$ ($M_{DM} < 20~{\rm GeV}$), would be indicative of a DM 
 candidate which is a relic from non-standard cosmology since the thermal  
 relic density is roughly two orders of magnitude higher than the observed 
 value of relic density.

\begin{figure}[htb!]
\begin{center}
\includegraphics[scale=0.25]{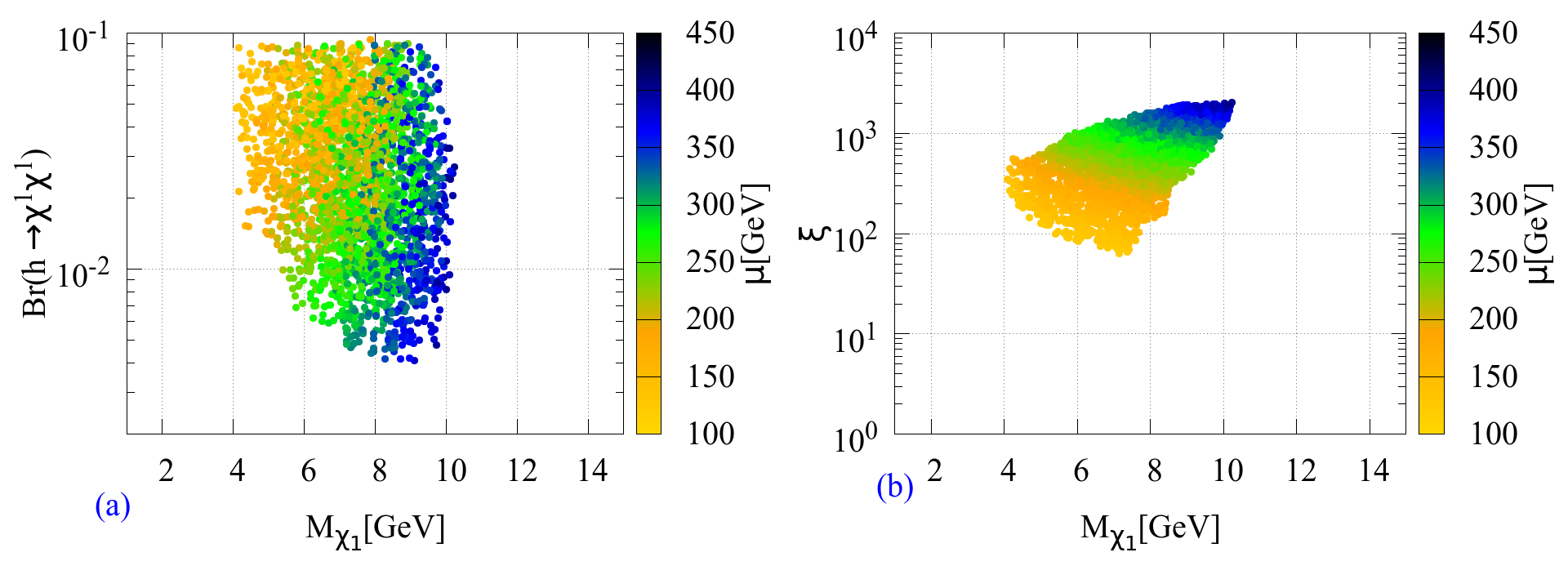}
\caption{(a) represents scatter plot in the $Br(h \to \lspone \lspone) - 
M_{\lspone}$ plane and (b) represents scatter plot in the $\xi - 
M_{\lspone}$ plane for parameter space points which can be probed by 
PICO-250 through the SD WIMP-proton interactions and also by ILC through 
the Higgs to invisible branching. The color palette corresponds to the 
value of the higgsino mass parameter ($\mu$).}
\label{comp_pico_ilc_cat}
\end{center}
\end{figure}

Before concluding this subsection, we consider one last category of 
parameter space points, which can be probed by PICO-250 through the SD 
WIMP-proton based interactions and by ILC, through the Higgs to invisible 
branching. All these parameter space points would also be sensitive to the 
electroweakino searches at ILC since $\mu\leq 500~{\rm GeV}$. We show 
these parameter space points in the $Br(h \to \lspone \lspone) - 
M_{\lspone}$ and $\xi - M_{\lspone}$ plane in Fig.~\ref{comp_pico_ilc_cat}
(a) and Fig.~\ref{comp_pico_ilc_cat}(b), respectively. The color palette 
represents the value of $\mu$. 
The parameter space points are confined within  $M_{\lspone} \le 10~{\rm 
GeV}$  and have $\xi \sim 70 - 2\times 10^{3}$, therefore an observation 
within this scenario would reflect that the DM candidate ($\lspone$) is an 
artefact of non-standard cosmology.

\subsection{Detection at Xenon-nT and LZ }
\begin{figure}[htb!]
\begin{center}
\includegraphics[scale=0.20]{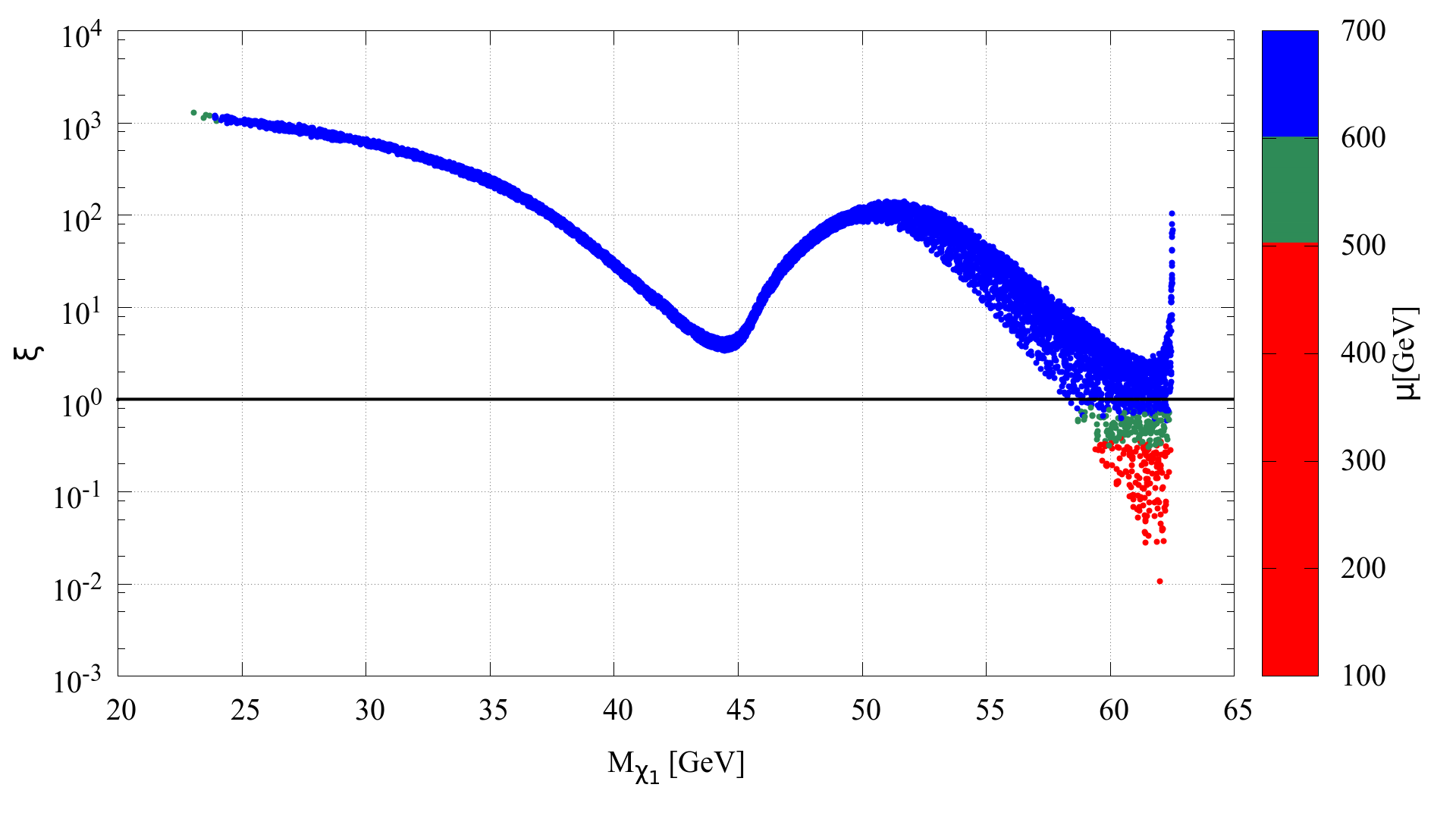}
\caption{Scatter plot in the $\xi - M_{\lspone}$ plane for parameter space 
points which can be probed by LZ through the SD WIMP-neutron interactions 
and by Xenon-nT through the SI WIMP-nucleon based interactions. The color 
palette corresponds to the value of the higgsino mass parameter ($\mu$).}
\label{comp_xent_lz_xi}
\end{center}
\end{figure}

The parameter space points which would be accessible to Xenon-nT through 
SI WIMP-nucleon based interactions and to LZ through SD WIMP-neutron based 
interactions only, have been shown in Fig.~\ref{comp_xent_lz_xi}, in the 
$\xi - M_{\lspone}$ plane. These parameter space points extend over 
$M_{\lspone} \sim 22 - 62.5~{\rm GeV}$, with $\xi$ varying between $0.02 - 
2000$. 
The parameter space can be split  into three different regions:  the non-
resonant region ($M_{\lspone} \lesssim 35~{\rm GeV}$) where $\xi \sim 300 
- 2000$,  the Z-resonance region ($M_{\lspone} \sim 35-55~{\rm GeV}$) with   
$\xi \sim 3 - 200$ and the Higgs resonance region ($M_{\lspone} \sim 
55-62.5~{\rm GeV}$), with $\xi$ varying between $0.02 - 60$. A signal in 
DD  and a rough determination 
of the DM mass in the first two regions would be indicative of the DM 
candidate being a relic from non-standard cosmology. On the other hand, 
the observation of a  DM mass in the Higgs resonance region would not be 
adequate enough to identify whether the DM candidate corresponds to the 
thermal picture or to a non-standard cosmology.

We observe a small set of parameter space points in the Higgs resonance 
region which are sensitive to the electroweakino searches at ILC as well, 
$\mu \leq 500 ~{\rm GeV}$, these points all correspond to  $\xi<1$. Adding 
the precise information on the value of $M_{\lspone}$ and $\mu$ from the 
ILC to an observation of a signal at Xenon-nT through the SI WIMP-nucleon 
based interactions, and at LZ, through the SD WIMP-neutron based 
interactions would indicate that the DM candidate ($\lspone$) is a 
standard cosmological relic.

\begin{figure}[htb!]
\begin{center}
\includegraphics[scale=0.20]{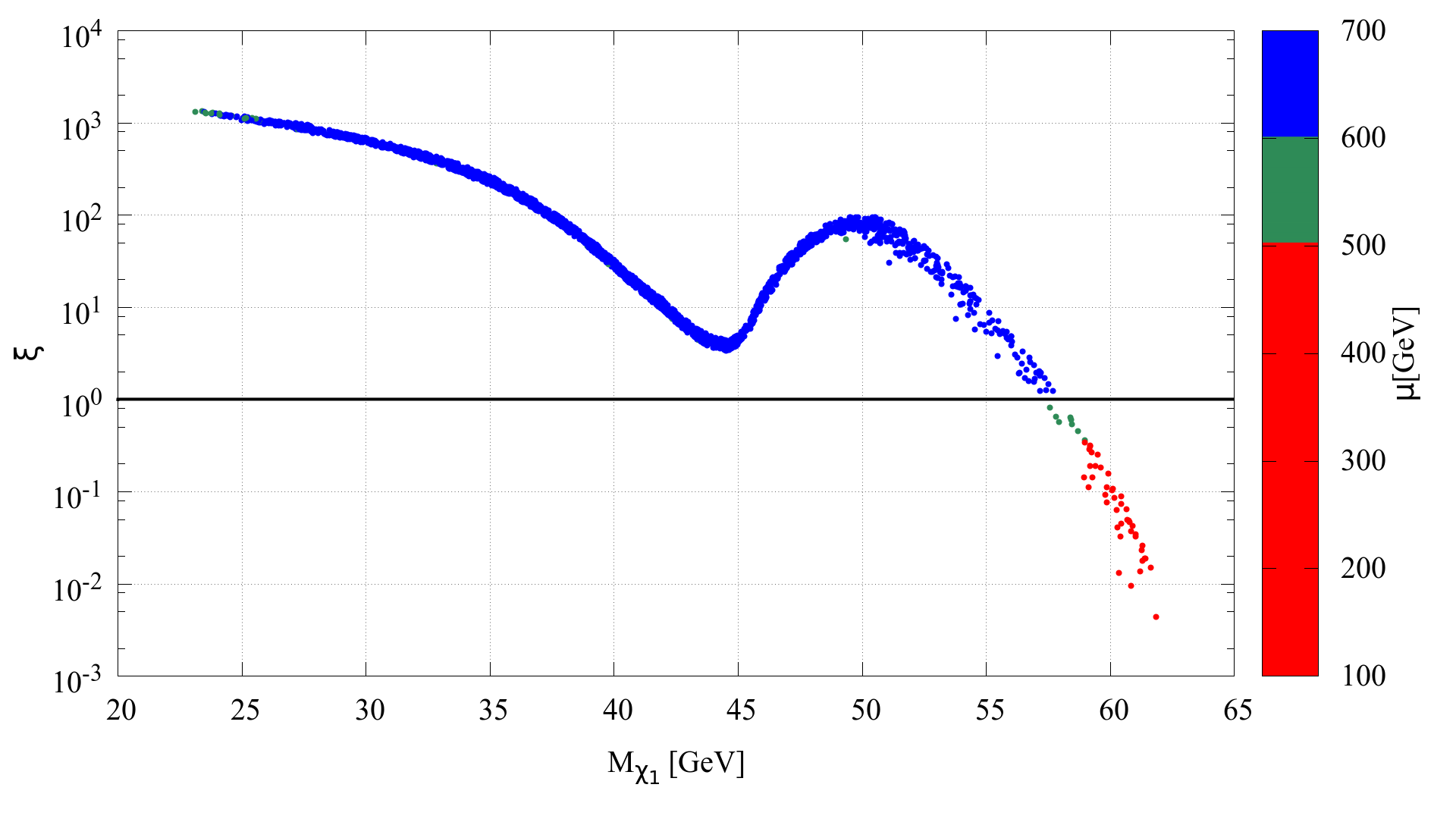}
\caption{Scatter plot in the $\xi - M_{\lspone}$ plane for parameter space 
points which can be probed by LZ through the SD WIMP-neutron interactions, 
by Xenon-nT through the SI WIMP-nucleon based interactions and by ILC 
through the Higgs to invisible branching fraction. The color palette 
corresponds to the value of the higgsino mass parameter ($\mu$).}
\label{comp_lz_xent_ilc_xi}
\end{center}
\end{figure}

Another similar subcategory of parameter space points are observed that 
would be detectable at Xenon-nT (through the SI WIMP-nucleon based 
interactions), at LZ  and at ILC (through the Higgs to invisible 
branching). These points extend over the three mass regions discussed 
above but are associated with a value for  $\xi$ lower by $\sim 1-2$ 
orders of magnitude. This is a result   of an increased coupling of the 
LSP to the Higgs and Z boson. Indeed, the majority of such parameter 
points have $\mu \leq 500~{\rm GeV}$ (shown in red color in 
Fig.~\ref{comp_lz_xent_ilc_xi}), therefore making them accessible to the 
electroweakino searches at ILC. With a very precise $M_{\lspone}$  and a 
DM signal 
at Xenon-nT (through SI WIMP-nucleon based interactions), at LZ (through 
SD WIMP-neutron based interactions) and at ILC (through both, 
electroweakino searches and Higgs to invisible branching fraction), one 
could conclude  that the DM requires a non-standard cosmology unless the 
LSP mass ($\lspone$) lies in the Higgs resonance region.

\subsection{Scenarios probed at Xenon-nT, PICO-250, LZ and ILC}   

\begin{figure}[htb!]
\begin{center}
\includegraphics[scale=0.20]{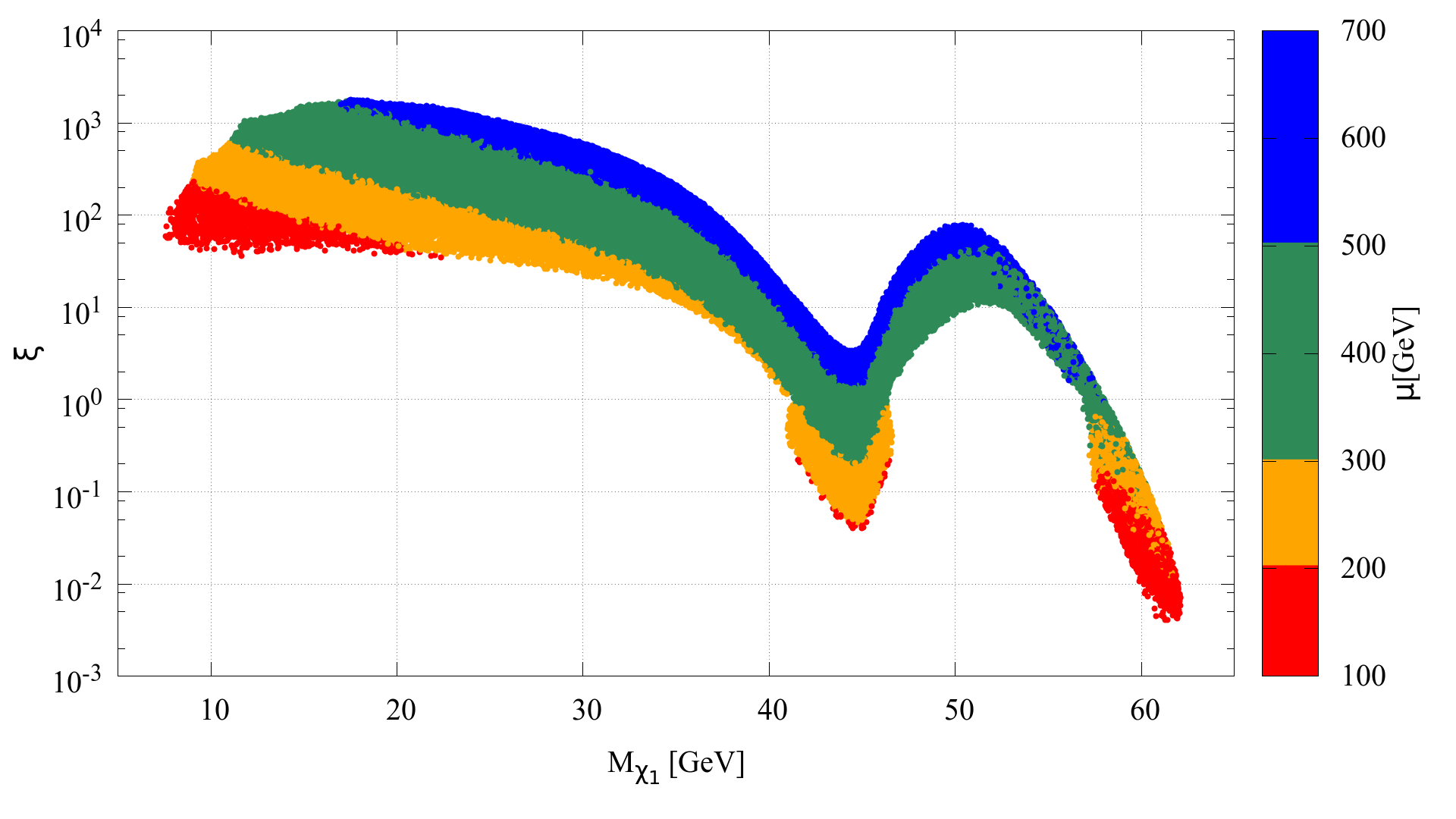}
\caption{Scatter plot in the $\xi - M_{\lspone}$ plane for parameter space 
points which can be probed by PICO-250 through the SD WIMP-proton 
interactions, by LZ through the SD WIMP-neutron interaction, by Xenon-nT 
through the SI WIMP-nucleon based interactions, and also by ILC through 
the Higgs to invisible branching fraction. The color palette corresponds 
to the value of the higgsino mass parameter ($\mu$).}
\label{comp_xent_pico_lz_ilc}
\end{center}
\end{figure}

In this subsection, we analyse those parameter space points which would be 
detectable at all the future direct detection experiments considered in 
this analysis, namely, Xenon-nT (through SI WIMP-nucleon based 
interactions), PICO-250 (through SD WIMP-proton based interactions), LZ 
(through the SD WIMP-neutron based interactions), and would also be 
accessible at ILC through the Higgs to invisible branching fraction and in 
some cases, through electroweakino searches. We show these parameter space 
points in Fig.~\ref{comp_xent_pico_lz_ilc} in the $\xi - M_{\lspone}$ 
plane. These parameter space points extend over a wide mass range 
$M_{\lspone}\sim 7-62.5~{\rm GeV}$, accompanied with a significant 
variation in $\xi$ as well, $\xi \sim 0.005 - 3000$.

The unique feature of this particular scenario is the presence of 
parameter space points in the Z resonance region ($m_{LSP}\approx 45$GeV) 
with relic density $\Omega h^{2} \leq 0.122$, which were absent in all 
other scenarios considered in this section. In addition, all such points 
have $\mu \leq 500~{\rm GeV}$ and are thus detectable through 
electroweakino searches at ILC as well. However, within the same Z 
resonance region, we also found parameter points with $\mu < 500~{\rm 
GeV}$ for  which $\xi>1$. As a result, it would not be possible to 
identify whether a signal in the Z resonance region is a signature of 
standard or non standard cosmology. Such a conclusion could however be 
reached for other mass ranges using the information on  $M_{\lspone}$ and 
$\mu$.

\begin{figure}[htb!]
\begin{center}
\includegraphics[scale=0.20]{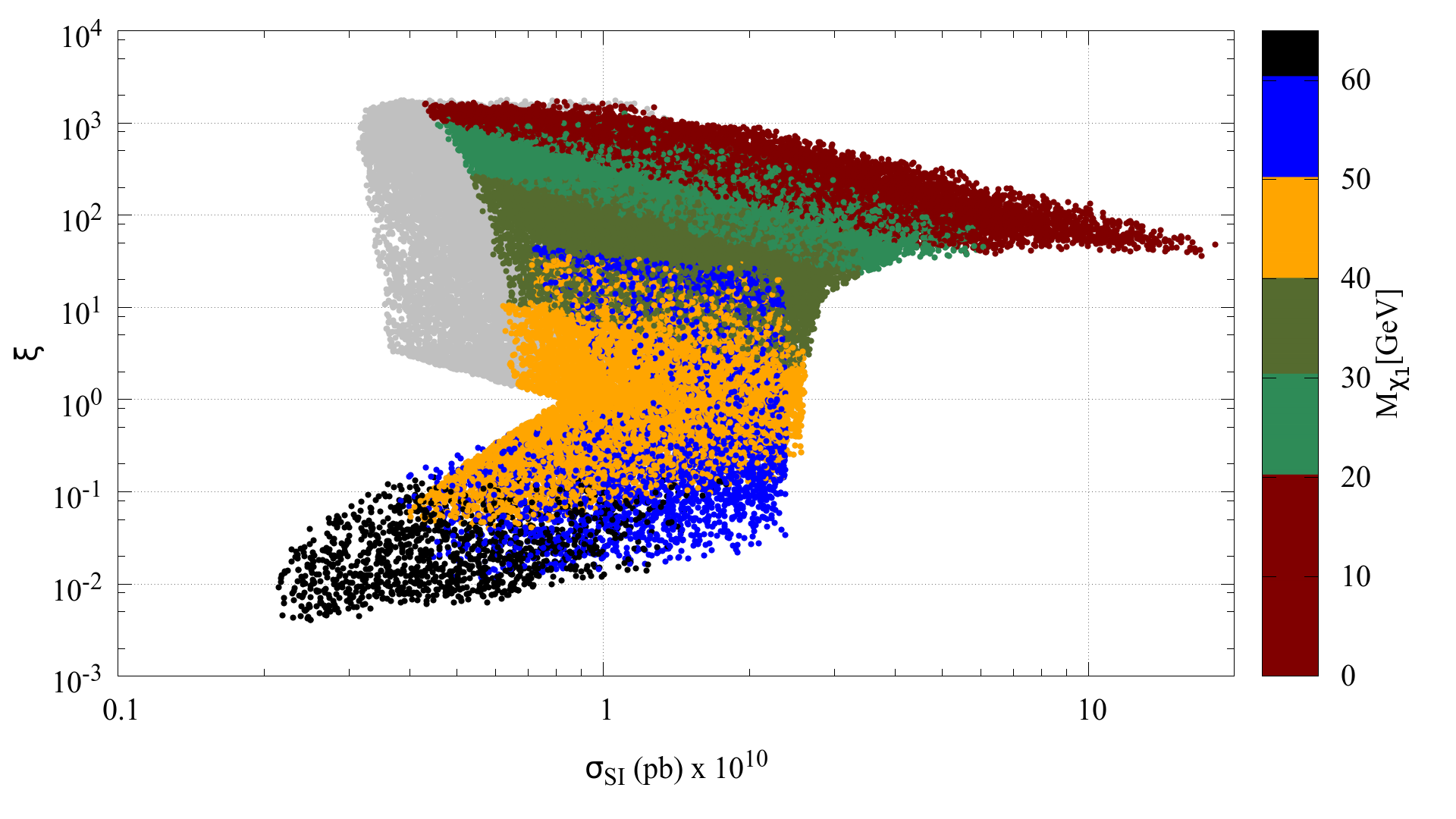}
\caption{Scatter plot in the $\sigma_{SI} - \xi$ plane for parameter space 
points which can be probed by PICO-250 through the SD WIMP-proton 
interactions, by LZ through the SD WIMP-neutron interaction, by Xenon-nT 
through the SI WIMP-nucleon based interactions, and also by ILC through 
the Higgs to invisible branching fraction. The color palette corresponds 
to the value of the LSP mass ($M_{\lspone}$).}
\label{comp_xent_pico_lz_ilc_si}
\end{center}
\end{figure}

In Fig.~\ref{comp_xent_pico_lz_ilc_si}, we show the parameter space points 
in the $\sigma_{SI} - \xi$ plane, with $M_{\lspone}$ represented through a 
color palette. The grey colored points correspond to those with $\mu > 
500~{\rm GeV}$ while those  represented through the color palette have  
$\mu \leq 500~{\rm GeV}$, making them accessible to ILC through the 
electroweakino searches as well. The additonal information from the 
determination of the LSP mass could be used to differentiate thermal and 
NSDM cosmological scenarios. 
 For example  parameter space points in the mass range $M_{\lspone} \leq 
 20~{\rm GeV}$ (shown in brown color) are indicative of the DM candidate 
 being an artefact of non-standard cosmology. Another interesting mass 
 region corresponds to $M_{\lspone} > 60~{\rm GeV}$ (shown in black 
 color), which are confined to the region with $\xi < 1$, and are hence, 
 indicative of a DM candidate which is a relic from standard cosmology.  
It is not possible to make similar arguments for the Z resonance region, 
where  parameter points with both, $\Omega h^{2} > 0.122 $ or  $\Omega 
h^{2} \leq 0.122 $ are found. We found that the additional information 
from $\sigma_{SD}$ measurements did not help to further  constrain the 
parameter space. 

\section{Prospects for high luminosity LHC}\label{sec:lhc}
In this section, we evaluate the role of the future runs of the LHC, with 
an integrated luminosity  up to 3000~fb$^{-1}$, in probing the light 
neutralino DM model.
We also briefly discuss certain unique signatures, which have a very 
negligible SM background, and could be used to obtain information on the 
gaugino sector of the MSSM  which would otherwise be difficult to access 
at ILC.

We begin by reminding ourselves of the following observation. It was seen  
in Sec.~\ref{sec:thermal} and Sec.~\ref{sec:non_standard} that the allowed 
parameter space was restricted to $Br(h \to \lspone \lspone)\lesssim 
10\%$, due to imposition of the Higgs signal strength constraints, derived 
by CMS and ATLAS through a combined analysis of the 
$7~\text{and}~8~\text{TeV}$ LHC data~\cite{Khachatryan:2016vau}. Hence the 
projected LHC limits on $Br(h \to invi.)$  mentioned in the introduction 
are not expected to imply any additional restriction on the allowed 
parameter space.
Future prospects of chargino neutralino searches have been studied in~ 
\cite{ATL-PHYS-PUB-2013-002}, in the context of a high luminosity LHC run 
($300~{\rm fb^{-1}}$, $3000~{\rm fb^{-1}}$ ). In \cite{ATL-PHYS-PUB-2013-002}, 
the LSP ($\lspone$) has been assumed to be bino-like, while 
, $\lsptwo$ and $~\chonepm$ have been assumed to be wino-like and 
degenerate in mass. Upper limits have been derived on the mass of 
$\lsptwo$ and $~\chonepm$ as a function of $M_{\lspone}$. For the 3 lepton 
final state from WZ-mediated simplified model, the exclusion contour goes 
upto $\sim 1110 ~{\rm GeV}$, while the $5\sigma$ discovery contour reaches  
$820~{\rm GeV}$, for $3000~{\rm fb^{-1}}$ of integrated luminosity (see 
Fig.~4 of \cite{ATL-PHYS-PUB-2013-002}). In the Wh simplified scenario, 
the exclusion contour reaches  $940~{\rm GeV}$, while the $5\sigma$ 
discovery contour reaches  $650~{\rm GeV}$ for $3000~{\rm fb^{-1}}$ of 
integrated luminosity (see Fig.~5(a) of \cite{ATL-PHYS-PUB-2013-002}).  
Thus the LHC will probe  models with a gaugino mass of the order of the 
TeV scale in the case $M_2 < \mu$.  Keeping these future projections in 
mind, we choose a representative benchmark point (BP 1) from the allowed 
parameter space with a chargino mass  within the projected LHC exclusion 
limits. However we do not demand the chargino to be wino-like. We 
intentionally choose BP 1 to be such a parameter space point which evades 
detection from all the future DD experiments considered in our analysis, 
namely, Xenon-nT, PICO-250 and LZ, and also has $Br(h \to \lspone \lspone) 
= 0.35\% $, making it inaccessible to the ILC through the Higgs to 
invisible searches. However, ino searches at ILC will be able to probe BP 
1 (since, $\mu = 442~{\rm GeV}$). BP 1 corresponds to the following set of 
input parameters:
\begin{eqnarray}
M_{1}~=~10.6~ {\rm GeV}, \quad M_{2}~=~812.6~ {\rm TeV}, \quad 
\tan{\beta}~=~42.8,\quad \mu ~=~442~ {\rm TeV}, \nonumber \\
 \quad  m_{\tilde{Q}_{3l}} =  8.42~{\rm TeV}, \quad m_{\tilde{t}_{R}} =  
 3.42~{\rm TeV}, \quad m_{\tilde{b}_{R}} =  4.93~{\rm TeV}, \quad  
 M_{3}~=~4.36~ {\rm TeV}, \nonumber \\
\quad A_{t}~=~2.42~ {\rm TeV}, \quad A_{b}~=~0~{\rm TeV}~=~A_{\tau} \nonumber \\
\quad m_{\tilde{Q}_{2l,1l}} =  3.00~{\rm TeV}, \quad m_{\tilde{c,u}_{R}} =  
3.00~{\rm TeV}, \quad m_{\tilde{s,d}_{R}} =  3.00~{\rm TeV}, \quad 
m_{slepton}~=~3~{\rm TeV} \nonumber \\  
\label{Parameter_space}
\end{eqnarray}

BP 1 is characterized by  a heavy wino component ($M_{2} = 812.2~{\rm 
GeV}$), resulting in nearly degenerate wino-type $\lspfour$ ($M_{\lspfour} 
= 822.81~{\rm GeV}$)  and $\chtwopm$ ($M_{\chtwopm} = 822.83~{\rm GeV}$). 
The LSP ($\lspone$) has a dominant bino fraction with mass 
$M_{\lspone}=10.3~{\rm GeV}$. The higgsino mass parameter ($\mu$) lies at 
an intermediate value of $442~{\rm GeV}$ resulting in 
$\lsptwo,~\lspthree,~\chonepm$ being dominantly higgsino-type with masses, 
$M_{\lsptwo}~=~436.2~{\rm GeV},~M_{\lspthree}~=~-446.4~{\rm 
GeV},~M_{\chonepm}~=~436.0~{\rm GeV}$ and thus accessible at the ILC. 
At the LHC, the production cross-section of winos is large and specific 
signatures can be found in
the cascade decay of the directly produced wino-type chargino/neutralino 
pairs, $ p p \to \lspfour \chtwopm$. 
The cascade decay of the wino-type $\lspfour, \chtwopm$ will be through 
the intermediate higgsino-type inos. One such final state topology would 
be the $ZZWh+\met$ final state, resulting from a cascade decay of the 
form, $\lspfour \to \lsptwo + Z,~\lsptwo \to \lspone + h$ and $\chtwop \to 
\chonep + Z,~\chonep \to \lspone + W^{+}$. Summing up all possible decay 
modes of the $\lspfour,\chtwopm$ pair, which end up in the $ZZWh+\met$ 
final state, we obtain a total branching of $\sim 15\%$. Considering the 
direct pair production cross-section, $\sigma(pp \to \lspfour \chtwopm) 
\sim 3 ~{\rm fb}$, evaluated using PROSPINO for $\sqrt{s}~=~14~{\rm 
TeV}$), we expect to produce $\sim 1350$ events in the channel  $pp \to 
\lspfour \chtwopm \to ZZWh+\met$ at  LHC for $3000~{\rm fb^{-1}}$ of 
integrated luminosity. Numerous other final states of the form $VVVW + 
\met$ or $VWWW + \met$ ($V = Z,h$), are also possible, all have a minimal 
SM background. For example, the $ZZWh+\met$ final state could be examined 
in the $5l + b\bar{b}+ \met$ channel, which has a negligible standard 
model background. The observation of a signal in these search channels, 
besides giving precious information on the hierarchy of the 
neutralino/chargino masses, would open the possiblity of obtaining  some 
rough estimate of $M_{2}$ from mass difference measurements.
In addition, the ILC will be able to perform very precise measurements of 
$\mu$  and $M_{\lspone}$ for BP1. For this benchmark, this information 
coupled with the non observation of light sfermions at the LHC and ILC 
will be sufficient to establish that the neutralino LSP cannot be a 
thermal DM candidate. In general, and especially for LSP with masses near 
$m_h/2$, 
the determination of only three parameters of the gaugino sector  is not 
sufficient  to establish whether the neutralino LSP observed is a thermal 
DM candidate or one needs to appeal to a non-standard cosmological model. 
In particular information on the fourth parameter, $\tan{\beta}$ is 
needed. In  favourable circumstances it could  be extracted   from  
pseudoscalar searches especailly if its value is large.  Recall that  the 
pseudoscalar production cross-section in the $b\bar{b}A$ mode is directly 
proportional to $({\tan{\beta})}^{2}$. A detailed investigation of relic 
density reconstruction  is beyond the scope of this work.

\section{Conclusion}\label{sec:conclusion}

We have revisited the case of the light neutralino DM in supersymmetry and 
investigate the impact of a precise measurement of the Higgs invisble 
width on the allowed parameter space of the pMSSM where only 
electroweakinos and the third generation fermions are allowed to be below 
2 TeV.
In the standard cosmological scenario where the neutralino is in thermal 
equilibrium with SM in the early universe, the light neutralino is 
confined  to two narrow range of masses around $m_Z/2$ and $m_h/2$.  Both 
region will be probed entirely by the Xenon-1T direct detection experiment 
while only the first region can be entirely probed  by a precise 
measurement of the Higgs invisible width achievable at the future ILC. 
Direct searches for higgsino at the ILC will allow to cover partly the 
Higgs funnel region. 
These conclusions are based on the assumption that there can be another DM 
component to explain the relic density when the neutralino DM is found to 
be underabundant. This approach is rather conservative since if we instead 
invoke a non-thermal mechanism to bring the low relic density scenarios 
within the PLANCK range,
constraints from direct detection become more severe and only a small 
fraction pass the constraint from LUX.

The picture changes completely once we relax the relic density constraint 
by assuming that some non standard mechanism enhances the total entropy 
density by the late 
decay of a field to SM fields. The allowed range of masses for the 
neutralino LSP extends to $\sim$ 1 GeV and it becomes much more difficult 
to cover the full parameter space with SI direct detection. SD experiments 
such as LZ and PICO-250 can in principle extend the reach, especially for 
DM masses below  10 GeV. However 
the expected cross-sections for such low masses of the LSP in this case 
can be  below the coherent neutrino scattering limit and therefore 
unreachable by these detectors. Therefore,
a precise measurement of the Higgs width is extremely important and in 
several cases provides the only handle on the LSP.

In the event of a discovery, the complementary measurements of Higgs 
invisible width or new particles at colliders and of DM direct detection 
can shed light not just on the nature of the DM but also on the 
cosmological scenario, in some cases pointing necessarily towards non-
standard mechanisms for DM production.  Various combinations of 
measurements 
can potentially point to a particular region of parameter space.  For 
example, observation of a signal at Xenon-nT, combined with the 
observation of the LSP and of higgsinos of mass below 500 GeV at the ILC, 
would imply that compatibility with the MSSM can only be accommodated for 
two precise mass regions, one with $M_{\chi} \approx 10~GeV$ and the other 
with $M_{\chi} \approx 60~GeV$. Moreover the first region can be 
consistent only with a non-thermal mechanism while the second can be 
compatible with a thermal relic.

We expect that the searches for electroweakinos at the LHC 13 TeV will 
contribute to constraining the parameter space but scenarios with 
$\mu,M_2$ at the TeV scale will remain out of reach. In the optimistic 
case, where we have winos lying below 800 GeV, and higgsinos lighter than 
the wino, there are spectacular signatures 
like 4V(=W/Z/h) + $\PMET$, which are background free, and can be readily 
observed at the high luminosity LHC. In this case a rough value of $M_{2}$ 
can potentially be extracted while 
a precise value  of $\mu$ and $M_{1}$ can be extracted  from higgsino 
searches at the ILC if $\mu$ is less than 500 GeV.  However even in the 
optimistic scenario, where  the above parameters are measured to some 
degree of accuracy, the value of the relic density can only 
be restricted within a certain range due to the lack of measurements of 
other parameters of the model.

In summary, measurements at the ILC and in direct detection measurements 
will provide the most important hints in determining the precise nature of 
the light neutralino, 
and could elucidate the cosmological nature of the light neutralino dark 
matter. 

\section*{Acknowledgments}

This work is supported by the ``Investissements d'avenir, Labex 
ENIGMASS'', by the French ANR,  Project DMAstro-LHC, ANR-12-BS05-006,  
by the  CNRS LIA (Laboratoire International Associ\'e) THEP (Theoretical 
High Energy Physics) and the INFRE-HEPNET 
(IndoFrench Network on High Energy Physics) of CEFIPRA/IFCPAR (Indo-French 
Centre for the Promotion of Advanced Research). 
The work of BB is supported by the Department of Science and Technology, 
Government of India, under the Grant Agreement number 
IFA13-PH-75 (INSPIRE Faculty Award). The work of RMG is supported by 
the Department of Science and Technology, India under Grant No. 
SR/S2/JCB-64/2007. RMG and BB acknowledge hospitality at LAPTh where part 
of this work was carried out. The work of DS is supported by the National 
Science Foundation under Grant PHY-1519045.

\providecommand{\href}[2]{#2}\begingroup\raggedright\endgroup


\end{document}